\documentclass[superscriptaddress,preprintnumbers,floatfix,showpacs,showkeys,twocolumn]{revtex4}

\usepackage{epsfig}
\usepackage{amscd}
\usepackage{amsmath}
\usepackage{amssymb}
\usepackage{graphicx}
\usepackage{dcolumn}
\usepackage{bm}
\usepackage{epsfig}
\usepackage{subfigure}
\usepackage{algorithmic}
\usepackage{algorithm}

\newtheorem{conjecture}{Conjecture}
\newtheorem{dataset}{Data set}

\newlength{\arrayrulewidthOriginal}
\newcommand{\Cline}[2]{%
  \noalign{\global\setlength{\arrayrulewidthOriginal}{\arrayrulewidth}}%
  \noalign{\global\setlength{\arrayrulewidth}{#1}}\cline{#2}%
  \noalign{\global\setlength{\arrayrulewidth}{\arrayrulewidthOriginal}}}

\begin{document}
\bibliographystyle{unsrt}

\title{Using time-delayed mutual information to discover and interpret temporal
  correlation structure in complex populations} 

\author{D. J. Albers}
\email{david.albers@dbmi.columbia.edu}
\affiliation{Department of Biomedical Informatics, Columbia
  University, 622 West 168th Street, VC-5, New York, NY 10032}

\author{George Hripcsak}
\email{hripcsak@columbia.edu}
\affiliation{Department of Biomedical Informatics, Columbia
  University, 622 West 168th Street, VC-5, New York, NY 10032}

\date{\today}

\begin{abstract}
  This paper addresses how to calculate and interpret the time-delayed
  mutual information for a complex, diversely and sparsely measured,
  possibly non-stationary \emph{population} of time-series of unknown
  composition and origin. The primary vehicle used for this analysis
  is a comparison between the time-delayed mutual information
  \emph{averaged} over the population and the time-delayed mutual
  information of an aggregated population (here aggregation implies
  the population is conjoined before any statistical estimates are
  implemented). Through the use of information theoretic tools, a
  sequence of practically implementable calculations are detailed that
  allow for the average and aggregate time-delayed mutual information
  to be interpreted. Moreover, these calculations can be also be used
  to understand the degree of homo- or heterogeneity present in the
  population. To demonstrate that the proposed methods can be used in
  nearly any situation, the methods are applied and demonstrated on
  the time series of glucose measurements from two different
  subpopulations of individuals from the Columbia University Medical
  Center electronic health record repository, revealing a picture of
  the composition of the population as well as physiological features.
\end{abstract}


\preprint{arxiv.org/abs/nlin/0607XXX}


\maketitle


\begin{quotation}
  \textit{In this paper we show how to apply time-delayed mutual information
  to a sparse, irregularly measured, complicated population of
  time-dependent data. At a fundamental level, the technical problem
  is a probability density function (PDF) estimation problem;
  specifically, one can \emph{average} PDF estimates \emph{or} one can
  aggregate the data set \emph{before} estimating the PDF. To
  understand and interpret these two means of coping with a population
  of time-series, one must address four issues: (i) estimator bias;
  (ii) normalization, or distribution support-based effects; (iii)
  deviations from the single source case for average and aggregate;
  and (iv) practical interpretation. Scientifically, this paper works
  to develop an infrastructure, and demonstrates how to use it, by
  studying the time-dependent correlation structure in physiological
  variables of humans --- in a population of glucose time-series. In
  the end, we not only provide a practically actionable set of
  information theoretic computations that yield insight into the
  population composition \emph{and} the time-dependent correlation
  structure, but we also detail the time-dependent correlation
  structure and the degree of homogeneity within a broad population of
  humans via their glucose measurements.}
\end{quotation}

\section{Introduction}
\label{sec:intro}

It is no surprise that aggregating collections of elements or data
streams can allow for a productive analysis and understanding of the
individual elements that make up the aggregated population. In fact,
the aggregation of many elements into a measurable population can be
pivotal in providing a means to study systems where the individual
elements are difficult, expensive, or dangerous to measure.  (Note
that by aggregation, we mean combining \emph{sets} of measurements in
such a way that they can be treated as a single \emph{set} of
measurements that can be analyzed.) That aggregation provides a basis
for analysis lies in the fact that the application of most statistical
methods, such as statistical averages, probability density estimates,
and techniques based on such fundamental methods (i.e., information
theory, ergodic theory, etc.), require large numbers of data
points. While some fields have gained much from the analysis of
aggregated populations of elements --- such as advances made in the
physical sciences with the advent of statistical mechanics --- many
fields have not been so fortunate. A primary source of difficulty with
aggregation in these less fortunate contexts lies in the fact that
fortune or ruin often depends on the ability to aggregate measured
elements such that statistical averages can be taken. Usually this
means one must have a population of elements whose statistical
properties being quantified are drawn from the same
distributions. This requirement presents two inextricable problems,
verifying that a population is homogeneous enough to produce
representative statistics when aggregated, and determining whether a
statistical analysis technique will yield the same outcome for the
average over the population and for the aggregated population. 


With these broad issues in mind, here we focus on applying
time-delayed mutual information to a population in an attempt to
understand \emph{the time-dependent nonlinear correlation between
  measurements, or the degree of predictability of measurements for
  members of a population}.  We wish to apply this, however, to a
system whose members may: (i) have differing numbers of measurements;
(ii) have too few measurements for probability densities (or any other
statistical quantities) to be estimated; (iii) be non-stationary; (iv)
have very diverse underlying probability distributions or statistical
states; and (v) may be measured in a highly irregular manner in
time. In short, this paper details how to apply and interpret
information theoretic analysis to a diversely measured, possibly
statistically diverse population that needs to be aggregated for the
information theoretic quantities to be calculable. Thus, this paper
complements and contrasts with the research such as is presented in
Ref. \cite{reconstruct_short_time_series_kurths} where dynamical
reconstruction of a uniformly measured stationary systems with short
time-series are the focus. The particular population we focus on in
this paper is a subpopulation of human beings who received care at the
Columbia University Medical Center (CUMC). The particular time-series
we are focusing on are clinical chemistry measurements (measurements
such as glucose, that detail physiological functioning of humans) for
this population.  Nevertheless, it is important to note that the
analysis presented is not limited to any particular population of
measurements.


\subsection{A reader's guide: the outline of this paper} 

Broadly this paper can be split into two main components. The first
component is primarily theoretical and includes: a background section
(\ref{sec:background}); a section about TDMI-specific estimator bias
(\ref{sec:the_bias_section}); a section focusing on how the TDMI for a
population can deviate from the TDMI of an individual stationary
source (\ref{sec:TDMI_pop_deviation}); and finally a section
explaining how to use the TDMI population calculations to characterize
diversity in a population (\ref{sec:tdmi_based_methods}). Second,
following the more theoretical sections, are the computational
sections including: a section explaining how to use the TDMI
population calculations to characterize diversity in a population
(\ref{sec:tdmi_based_methods}); a section proposing some non-TDMI
based metrics for evaluating population diversity that help verify the
TDMI-based methods (\ref{sec:non_TDMI_hetero_methods}); a section
summarizing the TDMI methodology explicitly
(\ref{sec:interpretation_summary}); and finally the data-based section
\ref{sec:quantitative_examples} demonstrating the
methodology. Regardless of intent, readers will need the to read the
introduction sections \ref{sec:intro}-\ref{sec:background} and the
summary.

\section{Motivating examples}
\label{sec:EHR_intro}

The theory-based motivation for this work is to devise a way to
calculate and interpret the time-delayed mutual information (TDMI)
\cite{sprott_book} \cite{kantz_book} in the context of a
\emph{population} of time-series that are both sampled irregularly and
are from (possibly) statistically distinct sources.  More concretely,
the motivation for this work comes from the desire to understand human
health dynamics (i.e., physiology, complex phenotype definitions such
as diseases, basic biology, etc) based on the constrains of real data
present in the electronic health record (EHR) repository at Columbia
University Medical Center (CUMC) (note, CUMC is affiliated with
NewYork-Presbyterian Hospital).  These data represents all the
information that doctors at CUMC collect; the CUMC EHR is one of the
oldest and most complete EHRs in the country, and thus represents the
type of data that future EHRs will likely contain. EHR data are of note
because EHRs contain most of the \emph{macroscopic}, biologically
based, data on humans in existence.

For instance, the CUMC EHR contains information regarding $2.5$
million patients over $20$ years and contains graphical images,
laboratory data, drug data, doctor and nurse notes, billing data, and
demographic data, most of which is highly dependent on time; moreover,
the \emph{amount} of data is growing exponentially. Despite the
quantity of data, EHR data can be difficult to use; in particular, EHR
data is characterized by: diverse irregular sampling, measurements
correlated to statistical state, nonstationarity, statistically
diverse population, very large populations with few measurements, and
very diverse data types. Nevertheless, if these data prove to be
useful for understanding human dynamics, a subject that is not
completely without controversy \cite{accuracy_med_records}
\cite{use_and_abuse_EHR} \cite{warfarin_ehr}, it may be possible: to
define complex diseases and other phenotypes (based on real,
population scale data); to understand how disease and treatment of
disease evolve in complex and interconnected ways
\cite{pnas_red_blood_cells_anemia} \cite{shudo_clinical_kineticsI}; to
define completeness of medical records; correlate drugs to side
effects and benefits; to monitor population-wide disease spread and
evolution; and to carry out many other practical applications that can
be gained from understanding population-wide human health and biology.
The approach upon which this work is based represents a radical
departure from the standard utilization of biomedical data; here the
data are studied using nonlinear physics methodology and has been
termed by some \cite{physics_of_living_things} as the physics of
living things.

Of course, another advantage of motivating the work in this paper with
a data set with complex properties is that it allows for the generalization of
the results to many other contexts whose data have a subset of the
complexities.  Outside of laboratory science, nearly all data sets are
difficult to control and have many of the same problems that EHR data
have. Thus, we claim that while we apply our analysis in the context
of human health and physiology, our methods can be easily
generalized to nearly all time-dependent contexts; e.g., astronomy
\cite{non_uniform_astro}, geology \cite{non_uniform_seismology},
climatology \cite{non_uniform_paleoclimatology}, and genetics
\cite{non_uniform_genetics}.

\section{Information theory background}
\label{sec:background}

Begin with time-series, $X=(x_1(t_1), \cdots, x_N(t_N))$ of real
numbers. Next, denote all of the \emph{pairs} of points in $X$
separated by a either index time, $\tau=i - j$ (where $i>j$ are
the indices of $t_i$ and $t_j$ respectively), or real time, $\delta t
= t_i - t_j$ (again assume $t_i>t_j$), by $X[\tau]$ or $X[\delta t]$
respectively. Note that $\tau$ is always an integer while $\delta t$
can take continuous real values. For this section we will limit the
discussion to $X[\tau]$, but note that the $X[\delta t]$ case
follows identically. Note that in this circumstance, $X[\tau]$ can be
used to approximate a \emph{joint} (two-dimensional) PDF; further,
note that the marginal distributions of $X[\tau]$ are approximated by
$X[\tau](1)=X(i)$ and $X[\tau](2)=X(i-\tau)$ respectively.

To estimate either the information entropy, or the TDMI for this
time-series \cite{kantz_book} \cite{sprott_book}, one must first
estimate various probability density functions (PDF)
\cite{allofstatistics}. In order to specify a PDF, one needs to both
specify the support of the PDF, $S$, and the PDF itself,
$p(X)$. Moreover, intuitively, the support of the PDF is the interval
over which the $x_i$'s lie, or, the support of the PDF of $X$ is
$S=[\min(X), \max(X)]$. However, when \emph{estimating} a PDF from
data, the support will always be collected in a series of bins; thus,
there also exists an \emph{abstract support}, $\mathcal{S}$, which
consists of the \emph{explicit} bins of the data used to estimate the
PDF \emph{disconnected from the values the bins are assigned
  externally}. Thus $\mathcal{S}$ does not explicitly represent
numbers in $X$; while this may seem like a strange point to make, the
difference between $S$ and $\mathcal{S}$ will be critical later in
this paper. Finally, note, we will always assume that PDFs in this
paper have \emph{compact support} \cite{loeve}.

Now, given the random variable $X$ and its associated PDF, $p(X)$, the
information entropy of a time series generated by $X$ is defined by:
\begin{equation}
h_I = - \int_S p(X) \log(p(X)) dx.
\end{equation}
Similarly, the TDMI is defined by:
\begin{align}
\label{equation:tdmi_first_def}
&I(X(i); X(i-\tau)) = \notag \\
&I(X[\tau]) = \\ \notag
&\int p(X(i), X(i-\tau)) \log
\frac{p(X(i), X(i-\tau))}{p(X(i)) p(X(i-\tau))} dX(i)dX(i-\tau)
\end{align}
Thus the TDMI can be thought of as an auto-information measure that
depends on a delay (e.g., $\tau$ or $\delta t$).

Given this infrastructure, fundamentally there are two ways of
conjoining a population: (i) averaging the TDMI for each member of the
population; and (ii) aggregating the population \emph{before} the PDFs
are estimated \emph{without} intermixing the members of the
population. As we will see, in the context of a heterogeneous
population, these two approaches will yield both differing numerical
results and differing interpretations.

Computationally it is important to note that we will employ both a KDE
estimator \cite{kde_matlab_I} \cite{kde_MI_estimators} \cite{KDE_IEEE}
and a standard histogram estimator for all PDF calculations. We
explicitly use the estimator developed in Ref. \cite{kde_matlab_I}
with a Gaussian kernel and a bandwidth of $100$; the histogram
estimator is of our own design and has a bandwidth of $20$.  The
results detailed in this paper are relatively insensitive to these
parameter settings (e.g., a $10 \%$ change in the bandwidth will not
produce a qualitatively different result). Moreover, in this paper we
will estimate the bias using the fixed point bias estimation technique
\cite{estimation_of_MI_EHR}, which amounts to various random
permutations of temporal ordering of the time-series used to generate
the PDFs and will be introduced in more detail in section
\ref{sec:RP_bias_estimates}. Finally, while this paper only addresses
the continuous case, the discrete case follows more or less
identically with integrals replaced by sums.

\subsection{Average TDMI}
\label{sec:average_MI_same_sum_average}

To formulate the average TDMI for a population, we begin by arguing
that the average mutual information of a vector of individuals (a
population) is the same as the average of the mutual informations of
each individual, if the individuals are independent. These cases
represent conjoining a population \emph{after the PDFs have been
  estimated}; in essence we are just arguing that taking an average
before or after the TDMI integration is performed does not affect the
resultant TDMI.

Assume all processes are stationary.  Define a
vector-valued process $X$, where $X(t) = [X_1(t), X_2(t), \cdots,
  X_N(t)]$; this leads to a the following definition of multivariate
mutual information:
\begin{align}
\label{equation:1}
        &I[X(t); X(t+j)] = \notag \\
&\int p(X(t), X(t+j)) \\ \notag
&\log \frac{p(X(t),X(t+j))}{p(X(t)) p(X(t+j))} dX(t) dX(t+j) 
\end{align}
noting that $p(\cdot)$ is the probability density associated with the
given random variable, and $X(\cdot)$ and $dX(\cdot)$ are both
vectors. We want the following statement to be true:
\begin{equation}
\label{equation:I_equality}
\frac{1}{N} I[X(t); X(t+j)] = \frac{1}{N} \sum_{i=1}^N I[X_i(t); X_i(t+j)]
\end{equation}
We claim that the \emph{sufficient} condition for
\ref{equation:I_equality} to hold is for the $X_i$ processes to be
non-interacting, or statistically independent.  It is important to
note that it is \emph{not} necessary that the $X_i$'s be
non-interacting copies of the \emph{same process} --- the processes
only have to be statistically independent. It is not too difficult to
verify our claim algebraically, one merely applies the chain rule for
mutual information to Eq. \ref{equation:I_equality}; moreover,
conceptually understanding why our claim is correct is rather
straightforward.  Begin by noting that if the $X_i$'s are independent,
they form an orthogonal set of probability densities, or a product
measure on $N$-dimensional Euclidean space.  Thus the integral of each
variable will be independent of the others simply because the
variables are orthogonal and thus not functions of one another (c.f.,
Fubini's theorem \cite{whezyg}).

The conclusion is that, the average TDMI for the population is simply
the \emph{canonically calcuated} TDMI for the individuals of the
population, averaged.

\subsection{Aggregate TDMI}

To understand the construction where the population is aggregated
\emph{before} the PDFs are estimated, assume, as we did in section
\ref{sec:average_MI_same_sum_average}, a stationary, vector-valued
process $X$, where $X(t) = [X_1(t), X_2(t), ... X_N(t)]$, where $N$
denotes the number of individuals in the population.  Next, assume
that each element emits a time-series of length $n_i$; without loss of
generality, in this section assume that $n_i=n$.

Aggregating the population into a time-series for which the PDFs can
be estimated can be done in one of two ways. The first method involves
concatenating the entire set of time-series into one scalar
time-series of length $N n$ and then treating this concatenated
time-series like a time-series from a single source; denote this
aggregation method as \emph{inter-source aggregation}.  We will
\emph{not} study this as this calculation needlessly adds noise via
the \emph{intermixing of elements} and is hard to rectify with
mathematics. The second method, denoted the \emph{intra-source
  aggregation} because sources are not intermixed within pairs of
points, involves explicitly collecting pairs of points restricted to
individuals.  Specifically, the pairs of points are chosen such that
the \emph{individual pairs of points always originate from the same
  individual}, and then these \emph{sets} of pairs of points are
conjoined such that the PDFs can be estimated. Thus, this method mixes
individuals by including pairs of points from many individuals, but
\emph{does not mix individuals by pairing points from differing
  individuals.}

To concretely specify what intra-source aggregation means, begin with
the time series:
\begin{equation}
(x_{1 1}, x_{1 2}, \cdots, x_{1 n}, x_{2 1}, \cdots, x_{N n})  
\end{equation}
where, given an $x_{i j}$, $i$ specifies the individual, $j$
specifies the time, and a time-delay of $\tau$ for which the TDMI is
to be calculated. The intra-source pairs that will be aggregated and
used for estimating the PDF are then:
\begin{align}
\label{equation:explicit_aggregated_set}
(x_{1,1}, &x_{1, \tau}) \notag \\
(x_{1,2}, &x_{1, 1+\tau}) \notag \\
&\vdots \notag \\
(x_{1,n-\tau}, &x_{1, n}) \notag \\
(x_{2,1} , &x_{2, \tau}) \\ \notag
&\vdots \\ \notag
(x_{2,n-\tau}, &x_{2, n}) \\ \notag
&\vdots \\ \notag
(x_{N,n-\tau}, &x_{N, n})
\end{align}
Thus, denote the \emph{left} column by $X_1^{n-\tau}$ and the
\emph{right} column by $X_{\tau}^{n}$.  Moreover, denote the TDMI
calculated between these two columns as
$I(X_1^{n-\tau}; X_{\tau}^{n})$.  

Much of the rest of this paper is dedicated to quantifying the
implications and interpretations for when, and conditions under which
the average and aggregate TDMIs differ. However, by comparing average
to aggregate TDMI we will also see that, very often (but not always),
the aggregate TDMI will form an upper bound on the TDMI of an individual.

\section{TDMI-specific estimator biases}
\label{sec:the_bias_section}

All statistical estimates have bias associated with them. Here we
focus on three sources of bias that are particular to the estimation
of the TDMI for a population: (i) sample-size-dependent estimator bias
effects for the average versus the aggregate TDMI; (ii) the basic
methodology we use for numerically estimating the bias for the TDMI
calculation; and (iii) a source of non-estimator bias that is
particular to the TDMI aggregation case --- a sort of filtering bias.

\subsection{Sample size dependent estimator bias effects}
\label{sec:estimator_bias}

A \emph{practical} reason why the \emph{order of aggregation} matters
for estimating probability densities lies in the fact that most
probability density estimation techniques have \emph{estimator bias}
that is, to first order, proportional to one over the number of points
to a power of at least one. Thus, because we are interested in coping
with populations of poorly measured individuals, and because we are
comparing two methods of conjoining those individuals, it is important
to understand how the number of data points will broadly affect
estimator bias in the average and aggregate TDMI calculations.


Begin with a more computationally minded definition of the TDMI for a
single time-series from a single source with $n$ points:
\begin{equation}
I[X_i(t); X_i(t-j)] = I_{X_i}(n) + B_{E}(n)
\end{equation}
where $I_{X_i}(n)$ is the estimated TDMI for the $n$ pairs of points
of $X$ and $B_{E}(n)$ is the total estimator bias of the calculation
with $n$ pairs of points.  Note that while explicit bias calculations
for the entropy and TDMI calculations can be found in
Refs. \cite{est_bias_histo_entropy}, \cite{est_bias_histo_MI}, and
\cite{estimation_of_MI_EHR}, it will suffice to notice that for most
PDF estimators (i.e., for kernel density estimators, or histogram
style estimators), the bias estimates will follow:
\begin{equation}
B_{E}(n) \sim n^{-1}
\end{equation}
Nevertheless, it is worth noting that there is also a
estimator-specific, bandwidth-specific factor on $B_{E}(n)$ that is
dependent on the \emph{proportion of support} (e.g., number of bins) for
which there exist no data points, and this factor can be important
when $n$ is small (c.f., \cite{est_bias_histo_MI} where this effect is
carefully quantified for the histogram estimator). To see how the bias
of averaging TDMI over the population versus the bias of the TDMI for
the pre-PDF-estimation aggregated populations differ, partition the
time-series of length $n$ into $m$ pieces, where $\frac{n}{m}$ is a
positive integer (thus, $m$ divides $n$ evenly and $n \geq m$).  Now,
consider the difference between $I[X_i(t), X_i(t-j)]$ calculated on
\emph{a single} time-series of length $n$ , and $I[X_i(t), X_i(t-j)]$
calculated on $m$ disjoint time-series of length $\frac{n}{m}$ and
then averaged.  More specifically, consider:
\begin{equation}
\label{equation:eq1}
I = I[X_i(t), X_i(t-j)] = I_{X_i}(n) + B_{E}(n)
\end{equation}
versus 
\begin{equation}
\label{equation:eq2}
I' = \frac{1}{m} \sum_{i=1}^m  I_{X_i}(n/m) + B_{E}(n/m)
\end{equation}  
Now, if the bias, $B_{E}$, scaled linearly in the number of points, $n$,
then the bias contribution of Eq. \ref{equation:eq1} will be the same
as the bias contribution of Eq. \ref{equation:eq2}.  However, we know
the bias obeys a power-law in the number of points, $n$, so we get the
difference between bias estimates to at least be:
\begin{align}
\delta B_{E} &= (\frac{1}{m} \sum_{i=1}^m B_{E}(n/m)) - B_{E}(n) \\
&\sim \frac{m-1}{n}
\end{align}
where $\delta B_{E} > 0$ for all $m>1$.  Or, said differently,  
\begin{equation}
\frac{1}{m} \sum_{i=1}^m B_{E}(n/m) \geq B_{E}(n)
\end{equation}
where equality is satisfied only when $m$ is one, or when the
population consists of a single element.  Note that when the
population is particularly poorly sampled, say one or two measurements
per element of the population, then $m \approx n$ and thus the
\emph{difference in the bias} of the population average versus the
aggregated population will be will be order one. More importantly,
\emph{averaging the MI of many poorly sampled individuals} will not
help the MI converge to its bias-free, high cardinality estimate.

Aside from the overall effect of $n$, there are other small sample
size effects, and these effects can have profoundly different outcomes
depending on the estimator. For instance, in the presence of few
points, a KDE estimator will often, in the name of smoothing,
over-estimate the probability for empty portions of the support,
resulting in a PDF estimate that is closer to a uniform random
variable. Thus, a KDE-PDF based TDMI calculation will likely
\emph{underestimate} the TDMI. In contrast, a histogram estimator will
underestimate the probability for empty portions of the support, thus
yielding a more sharply peaked distribution that will yield an
over-estimate of the TDMI. Because of these opposing effects, it is
possible to verify the existence of finite-size effects by simply
observing the difference between the KDE and histogram estimated TDMI
estimates for the same data set.

In the end, because we are working to understand how to estimate the
TDMI in the context of large, poorly measured populations, there will
be a significant advantage to aggregating populations \emph{before}
estimating the PDFs necessary to carry out the TDMI calculations from
the perspective of estimator bias minimization.

\subsection{Fixed point bias estimate for average and aggregate populations}
\label{sec:RP_bias_estimates}

The fixed point TDMI bias estimation method
\cite{estimation_of_MI_EHR} attempts to estimate the $\tau = \infty$
TDMI by randomly permuting the time-ordering of one of the sets of
pairs used to estimate the distributions for a given $\delta t$ or
$\tau$. Fundamentally, there are two different methods for estimating
the TDMI fixed point (if it exists), random permutation \emph{within
  the individuals} (i.e., not mixing individuals), and random
permutation \emph{over the entire population}, thus intermixing
individuals.

The first method, \emph{individual-wise random permutation} (IRP),
involves randomly permuting the temporal ordering of one column
(without replacement) of the data set used to estimate the
distributions \emph{without intermixing individuals}, or:
\begin{equation}
\label{equation:B_IRP}
  B_{IRP}(\tau, n)=\lim_{Z \rightarrow \infty} \frac{1}{Z} \sum_{i=1}^Z I(X_1^{n-\tau}, \mathcal{X}_{\tau}^{n}(i,t))
\end{equation}
where $\mathcal{X}_n^{\tau}(i,t)$ is the $i^{th}$ random permutation
(without replacement) of the \emph{left index} of the column vector
$X_{\tau}^n$ (i.e., do not permute the first index of $x_{i,j}$ from
equation \ref{equation:explicit_aggregated_set}). The IRP-method
random permutation occurs \emph{only within an individual} and not
across the population, thus destroying information about only
time-based correlations while preserving inter-individual
information. Finally, there will exist a IRP bias estimate for both
the average and aggregate TDMI cases, $\bar{B}_{IRP}$ where
Eq. \ref{equation:B_IRP} is specified for a single individual and then
averaged over the population, and $\hat{B}_{IRP}$ which is specified
exactly as per Eq. \ref{equation:B_IRP}.

The second method, \emph{population-wide random permutation} (PRP),
which exists only in the aggregated population context, involves
randomly permuting, without regard to the individual, one column of
the \emph{entire populations'} data set used to estimate the PDFs or,
\begin{equation}
  \hat{B}_{PRP}(\tau, n)=\lim_{Z \rightarrow \infty} \frac{1}{Z} \sum_{i=1}^Z
  I(X_1^{n-\tau}, \mathcal{X}_{\tau}^{n}(i,N,t))
\end{equation}
where $\mathcal{X}_n^{\tau}(i,N,t)$ is the $i^{th}$, random
permutation (without replacement) of the \emph{both indices} of column
vector $X_{\tau}^n$. Because the PRP estimate intermixes both the
population and time, the PRP destroys information about \emph{both}
intra-individual time correlations \emph{and} inter-individual
information (i.e., information about differences in normalization or
the supports). In the context of a single source, $\bar{B}_{IRP} =
\hat{B}_{IRP}(n)=\hat{B}_{PRP}(n)$. Similarly, when the population is
both relatively uniform over both the PDFs and the support of the
PDFs, then the PRP bias estimate will be equivalent to the bias
estimate of the IRP, and thus can be thought of as an estimate of the
estimator bias.  However, if the support of the PDFs over the
population is not uniform (i.e., if the support of any of the
individuals of the population differs from the support of the
population), then the PRP bias estimate will differ from the IRP bias
estimate (we will discuss this explicitly in section
\ref{sec:support_effects_interp}). Note that $\bar{B}_{IRP}$,
$\hat{B}_{IRP}$, and $\hat{B}_{PRP}$ are dependent on both $\tau$ or
$\delta t$ (because of the filtering effect discussed in the next
section) and $n$, the number of points used in the estimate. In
general, we will drop the $n$ from the notation, and when there is not
a $\tau$ or $\delta t$ dependence, we will not include it in the
notation (in general, for the data sets and $\delta t$'s we consider
in this paper, there is not a strong $\delta t$ dependence).

\subsection{Non-estimator bias: how the TDMI calculation can act as a
  population filter}
\label{sec:filter_as_bias}

While it is clear that the TDMI calculation only applies to the data
used to estimate the PDFs, it is less obvious that \emph{the act of
  constructing the data sets used to estimate the PDFs} can
\emph{filter out substantial portions of the overall population.}
Specifically, because construction of the data sets for the PDF
estimation involves collecting all \emph{pairs} of points separated by
some time $\tau$ or $\delta t$, if some individuals do not have
\emph{pairs} of points separated by $\tau$ or $\delta t$, those
individuals will be filtered out of, or excluded from, the data set
used to estimate the PDFs and thus the TDMI. In this sense, the TDMI
calculation \emph{implicitly filters the population by measurement
  frequency}; this is not an externally imposed data constraint, it is
simply a result of calculating the TDMI in the context of population
whose elements do not have identical measuring frequencies.

To understand how this filtering bias can affect the results, consider
a polarized example population made up of two differently measured
subsets of individuals. Specifically, the first subset of the
population has individuals sampled once an hour for a month and the
second subset of the population has individuals sampled once a month
for $20$ years. These two population represent patients with acute and
chronic conditions, respectively. If the TDMI of the population is
calculated for any $\delta t$ less than a month, only data set one
will be represented. Similarly, if the TDMI is calculated for $\delta
t$ of a month or greater, only data set two will be represented. When
plotting the TDMI graph versus $\delta t$, the graph has, in a sense,
a bias.  Namely, two the graph represents two disjoint populations for
$\delta t >$ one month.

Of course, for real EHR data, even more complicated problems can
appear when the same individual is sampled at different rates
\emph{depending on the statistical state of the individual} (e.g., a
patient with a chronic and acute condition). This problem is
particularly acute for health care data because health correlates with
presence of measurement --- healthy patients are not measured often
while sick patients are --- thus leading to the possibility of having
different subpopulations or statistical states being filtered out when
calculating the TDMI for some $\delta t$ values.


Thus, when estimating a TDMI for a population, it is important to
quantify both who is populating the data set \emph{explicitly used} to
estimate the PDFs \emph{and} how the proportionality of the
subpopulations changes in the set used to estimate the PDFs as the
delay is changed. If the population and proportionality of
subpopulations in all the $\delta t$ or $\tau$ TDMI estimates does not
change, then the bias estimates are \emph{independent of the delay.}

\subsubsection{Methods for assessing $\delta t$ bin compositions}

To quantify the composition of the data set, begin with the following
notation: (i) $b_i(\tau)$ represents the number of pairs of points in
the $\tau$ time bin contributed by individual $i$; (ii)
$b_{max}(\tau)=N_{max}$ and $b_{min}(\tau)=N_{min}$ correspond to the
maximum and minimum number of pairs of points, over all individuals,
present in the data set; $N_{*}$ represents the sum of $b_i(\tau)$, or
the total number of pairs of points in the data set; (iii) $N$
represents the total number of individuals in the population; and
(iv), $\varsigma(\tau)$ represents the set of indices of individuals
monotonically ordered by increasing $b_i$. Based on these quantities,
define the following functions:
\begin{align}
\Theta(\varsigma(\tau)) &= b(\varsigma), \\
\tilde{\Theta}(\tilde{\varsigma}(\tau)) &= \frac{b(\frac{\varsigma (\tau)}{M})}{b_{max}}
\end{align}
noting that $\tilde{\Theta}(\tau)$ \footnote{It may seem odd to normalize
  indices, but this just keeps the domain of $\tilde{\Theta}$ between
  zero and one.} is $\Theta (\tau)$ normalized to lie on the unit
square. Next, define the following integral that quantifies the
population composition of the data set:
\begin{equation}
H_{\tilde{\Theta}} (\tau) = \int_{\tilde{\varsigma}} \tilde{\Theta} d \tilde{\varsigma}.
\end{equation}
When the time series of the members of the population are both
uniformly sampled and of the same length, $H_{\tilde{\Theta}} (\tau)$ will be
equal to one; thus the closer $H_{\tilde{\Theta}} (\tau)$ is to one, the more
composition of the data set includes the entire population uniformly,
while the closer $H_{\tilde{\Theta}} (\tau)$ is to zero, the more composition
of the data set represents a small subset of the population (possibly
only an individual). A second, more gross quantification of how the
population is represented in TDMI data set at a fixed $\delta t$ is
the percentage of individuals that contribute at least one pair to the
data set, or:
\begin{equation}
H_{b_i \neq 0} (\tau) = \frac{\# (b_i \neq 0)}{N}
\end{equation}
Note that an alternative, highly related quantity we have found useful
is the cumulative distribution function (CDF) of the $b_i$'s.

Finally, while it is tempting to think of the population makeup of the
$\tau$ data set as a measure of homogeneity within a population, this
interpretation is sometimes, but not always, correct. What
$H_{\tilde{\Theta}} (\tau)$, $H_{b_i \neq 0} (\tau)$, or any other
like-minded metric really detail is how a population is measured and
thus represented in a given $\tau$ or $\delta t$ bin. Specifically,
when \emph{measurement frequency} is correlated with statistical state
or dynamics, then it is likely that $\tau$ bins will filter a
population and make it more homogeneous. However, it is easy to think
of examples where \emph{measurement frequency} is random, or uncoupled
from a statistical state or dynamics, and in this case, all the
diversity of the population will be present in any given $\tau$ time
bin.

\section{Population-based deviations from the individual TDMI estimates}
\label{sec:TDMI_pop_deviation}

\subsection{Heterogeneity-based deviations from the individual: average
TDMI case}

To understand how representative the \emph{average MI} over the
population is of an individual in the population, begin by setting
$p_1$ as the PDF that most resembles the \emph{average} (choosing
$p_1$ to be the median among the $p_i$'s would work as well) among the
set of $p_i$'s \emph{relative to the abstract support},
$\mathcal{S}$; note that the average PDF is defined by:
\begin{equation}
\bar{p} = \frac{1}{N} \sum_{i=1}^N p(X_i[\tau]).
\end{equation}
Note that in this situation, every $p_i$ has the same
abstract support (by definition), which we will denote as
$\bar{\mathcal{S}}$. Further, note that it is possible to have a set
of $p_i$'s such that no $p_i$ resembles the mean \emph{graph} of the
$p_i$'s. Next, relative to $p_1$ we can now relate each $p_i$ to $p_1$
as follows:
\begin{equation}
p_i = p_1(\bar{\mathcal{S}}) - \bar{\epsilon}_i(\bar{\mathcal{S}})
\end{equation}
where $\bar{\epsilon}_i(\bar{\mathcal{S}})$ is distance between the
\emph{graphs} of $p_1$ and $p_i$ at a given value in
$\bar{\mathcal{S}}$. Recalling the definition of the TDMI, we get:
\begin{align}
\label{equation:ibar}
&I[X(t); X(t+\tau)] = \\ \notag
&\bar{I}(\tau) = \frac{1}{N} \sum_{i=1}^N \int p(X_i(j),
  X_i(j-\tau)) \\ \notag
&\log
  (\frac{p(X_i(j),X_i(j-\tau))}{p(X_i(j))p(X_i(j-\tau))} ) dX_i(t)
  dX_i(t+\tau) \\ \notag
&= \int \bar{\iota}(\tau) dX(t) dX(t+\tau).
\end{align}
Now, because integration is a linear operation, focus on the integrand
instead, or more specifically, focus on:
\begin{align}
&\bar{\iota}(\tau) = \\ \notag
&\frac{1}{N} \sum_{i=1}^N p(X_i(j),
  X_i(j-\tau)) \log
  (\frac{p(X_i(j),X_i(j-\tau))}{p(X_i(j))p(X_i(j-\tau))} ) \\ \notag
&= p(X_1(j), X_1(j-\tau)) \log
(\frac{p(X_1(j),X_1(j-\tau))}{p(X_1(j))p(X_1(j-\tau))} ) \\ \notag
&+ \bar{G}(N,
\epsilon_i, p(X_1(j),X_1(j-\tau)),p(X_1(j)),p(X_1(j-\tau)) ) \\ \notag
&= \bar{\rho}(\tau) + \bar{G}(\tau)
\end{align}
where, $\bar{G}(\tau)$ is given by:
\begin{equation}
\label{equation:gbar}
\begin{split}
\bar{G}(\tau) &= \\
&- \frac{1}{N} [ \sum_{i=1}^{N-1}
\left( \frac{\bar{\epsilon}_i}{p(X_1(j), X_1(j-\tau))} \right) \\
&\left(\log \frac{p(X_1(j),
  X_1(j-\tau))}{p(X_1(j))p( X_1(j-\tau))} \right) \\
&+ \log \left(
\frac{1-\frac{\bar{\epsilon}_i}{p(X_1(j),X_1(j-\tau))}}{(1-\frac{\bar{\epsilon}_i}{p(X_1(j))})
  (1-\frac{\bar{\epsilon}_i}{p(X_1(j-\tau))} )} \right) \\
&\left( \frac{\bar{\epsilon}_i}{p(X_1(j), X_1(j-\tau))} - 1 \right) ]
\end{split}
\end{equation}
(for a more explicit calculation of $\bar{I}$, c.f., appendix
\ref{sec:app1.1}). As each $\bar{\epsilon}_i$ goes to zero, $\bar{G}$
goes to zero; thus the more \emph{support independent} variance
(recall $\bar{\epsilon}_i$ is relative to the abstract support
$\bar{\mathcal{S}}$) there is within the population, the larger
$\bar{G}$ will be, and the less $\bar{I}(\tau)$ will represent the
TDMI of an individual element within the population. Written
explicitly, $\bar{I}(\tau)$ represents the ``average'' individual
\emph{plus} the \emph{sum of the deviations from that individual.}

\subsubsection{Entropy of the averaged population}

While the primary topic in this paper is the TDMI, we will contend
briefly with the TDMI for $\tau=0$, or the auto information. Based on
an identical means of calculation, the information entropy of a
time series for a population can be defined as follows:
\begin{align}
\bar{h}_I = -  \frac{1}{N} \int [ p_1 \log (p_1) &+ p_1 \sum_{i=1}^{N-1} \log (p_1
- \bar{\epsilon}_i) \\ \notag
&- \sum_{i=1}^{N-1} \epsilon_i \log (p_1 - \bar{\epsilon}_i) ] dx
\end{align}
Thus, when $\bar{\epsilon}_i \rightarrow 0$, the $h_I$ for the population
\emph{relative to the abstract support} tends toward the information
contained in an individual.

\subsection{Heterogeneity-based deviations from the individual: aggregate
TDMI case}

To understand how the diversity in the population is rendered via the
TDMI of the \emph{aggregated population} begin by recalling that the
TDMI for the aggregated set is defined by:
\begin{align}
\label{equation:aggregated_MI_def}
\hat{I}(\tau) =& I(X_1^{n-\tau};X_{\tau}^{n}) \\ \notag
 =& \int p(X_1^{n-\tau}; X_{\tau}^{n}) \log
 (\frac{p(X_1^{n-\tau}; X_{\tau}^{n})}{p(X_1^{n-\tau})
   p(X_{\tau}^{n})})dX_1^{n-\tau} dX_{\tau}^{n} \\ \notag
=& \int \hat{\iota}(\tau) dX_1^{n-\tau} dX_{\tau}^{n}
\end{align}
where, under ideal (single, stationary source) circumstances the PDF
of the aggregated density obeys
\begin{equation}
\label{equation:aggregation_is_sum_PDFs}
\hat{p}(X_1^{n-\tau}; X_{\tau}^{n}) = \frac{1}{N} \sum_{i=1}^N p(X_1^{n-\tau}(i); X_{\tau}^{n}(i))
\end{equation}
where $X_1^{n-\tau}(i)$ and $X_{\tau}^{n}(i)$ represent the PDF
restricted to individual $i$. Intuitively,
Eq. \ref{equation:aggregation_is_sum_PDFs} just says that we are
creating the aggregate PDF by summing the \emph{graphs} of all the
individuals \emph{relative to the union of the supports of all the
  individuals}, that is, relative to $\hat{S}=\cup_{i=1}^N \hat{S}_i$.

To choose a PDF that most closely resembles a centroid, it is helpful
to have a concept of abstract support; however, because $\hat{I}(\tau)$ is
defined relative to the actual support of the \emph{population}, the
individual population PDFs do not separate as naturally as in the
$\bar{I}(\tau)$ case. Nevertheless, conceptually, to define an abstract
support in the aggregate circumstance, one needs to, in spirit, construct
a situation where all the PDFs have roughly the same range or support.
There are several ways one can imagine achieving such goal; here will
define the \emph{abstract support}, $\hat{\mathcal{S}}$, such that
every patient has been renormalized to have the identical support ---
the unit interval (i.e., $[0,1]$). It is important to realize that
relative to the aggregate case there can be
a very severe difference between the TDMI of an aggregated population
defined on support of the $\hat{S}$ versus the abstract support
$\hat{\mathcal{S}}$. To allow for quantifying these potential
differences, define the TDMI for an aggregated population relative to
the abstract support, $\hat{\mathcal{I}}(\tau)$. Now, using the abstract
support, select $p_1$ in the same way we selected $p_1$ in the
previous section, by selecting the PDF that most closely represents
the mean over the population of PDFs \emph{relative to the abstract
  support}. This definition implies an important difference in how
$p_i$ is specified in the aggregate case versus the average case
because, despite the fact that we use an abstract support to select a
$p_1$, $\hat{I}(\tau)$ is \emph{not} calculated relative to the abstract
support, and thus the differences between $p_1$ and $p_i$ are instead
defined by:
\begin{equation}
\label{equation:p_hat_substitution}
p_i = p_1(\hat{S}) - \hat{\epsilon}_i(\hat{S})
\end{equation}
where $\hat{\epsilon}_i(\hat{S})$ is distance between the
\emph{graphs} of $p_1$ and $p_i$ at a given value in \emph{total
  support}, $\hat{S}$. Next, focusing on the integrand, $\hat{\iota}$,
and substituting Eq. \ref{equation:p_hat_substitution} into
Eq. \ref{equation:aggregation_is_sum_PDFs} and recalculating
$\hat{\iota}$ we arrive at (dropping the subscript on $p_1$):
\begin{align}
  &\hat{\iota}(\tau) = p(X_1^{n-\tau}; X_{\tau}^{n}) \log
  (\frac{p(X_1^{n-\tau}; X_{\tau}^{n})}{p(X_1^{n-\tau})
    p(X_{\tau}^{n})}) \\ \notag &+ \hat{G}(\tau)(N, \hat{\epsilon}_i,
  p(X_1(j),X_1(j-\tau)),p(X_1(j)),p(X_1(j-\tau))) \\ \notag &=
  \hat{\rho}(\tau) + \hat{G}(\tau)
\end{align}
where $\hat{G}(\tau)$ is explicitly given by:
\begin{equation} 
\label{equation:ghat}
\begin{split}
\hat{G}(\tau) =& \log \left(  \frac{1 - \frac{\sum_{i=1}^{N-1}
    \hat{\epsilon}_i}{N p(X_1^{n-\tau},X_{\tau}^{n})}}{(1 -
  \frac{\sum_{i=1}^{N-1} \hat{\epsilon}_i}{N p(X_1^{n-\tau})})( 1 - \frac{\sum_{i=1}^{N-1}
    \hat{\epsilon}_i}{N p(X_{\tau}^{n})})}   \right) \\
&\left(
p(X_1^{n-\tau},X_{\tau}^{n}) - \frac{\sum_{i=1}^{N-1}
  \hat{\epsilon}_i}{N}  \right) \\
-& \frac{\sum_{i-1}^{N-1}\hat{\epsilon}_i}{N} \log \left(  \frac{p(X_1^{n-\tau}, X_{\tau}^{n})}{p(X_1^{n-\tau})
   p(X_{\tau}^{n})} \right)
\end{split}
\end{equation}
(for a more explicit calculation of $\hat{G}$ and $\hat{I}$, c.f.,
appendix \ref{sec:app1.2}). Thus, as the \emph{average of the}
$\hat{\epsilon}_i$'s go to zero, $\hat{G}(\tau)$ will go to zero;
moreover, when \emph{both} the width of the band of PDFs decreases and
when the supports of the distributions overlap (i.e., when
$\cap_{i=1}^N \hat{S}_i \rightarrow \cup_{i=1}^N \hat{S}_i$), the TDMI
of the aggregate population ($\hat{I}$) will represent an individual within a
homogeneous population (because the individuals within the population
are similar). Similarly, when either the width of the band of PDFs
increases or the supports of the distributions becomes disjoint, (i.e.,
when $\cap_{i=1}^N \hat{S}_i \rightarrow 0$), $\hat{I}(\tau)$ will
represent the TDMI within the diverse population. Or, said
differently, the TDMI for the aggregated population will represent the
TDMI of the \emph{population} plus the \emph{sum of the individual
  based differences from the population.} As we will see in the
sections that follow, this second circumstance can lead to subtle
difficulties in interpretation. Finally, note that the calculation
that yielded $\hat{\iota}$ does not explicitly depend on the support;
the explicit $\hat{\epsilon}$'s will differ between $\hat{I}(\tau)$ and
$\hat{\mathcal{I}}(\tau)$, but the explicit form of $\hat{\iota}$ will not.

\subsubsection{Entropy of the aggregated population}

Again, while the TDMI is the primary topic of this paper, in both the
interest of completeness and later analysis, we define $h_I$ for the
aggregated population, which was calculated in analog with $\hat{I}$,
as follows:
\begin{align}
\hat{h}_I = -&\int p \log ( p - \frac{\sum_{i=1}^{N-1} \hat{\epsilon}_i}{N})
\\ \notag
&-\frac{\sum_{i=1}^{N-1} \hat{\epsilon}_i}{N} \log (p - \frac{\sum_{i=1}^{N-1}
  \hat{\epsilon}_i}{N}) 
\end{align}
In contrast to the situation where the information entropy is averaged
over the population, when the average $\frac{\hat{\epsilon}_i}{N}
\rightarrow 0$, the information entropy for the aggregated population,
$\hat{h}_I$, \emph{relative to the real support of the population}
tends toward the information contained in an individual who has the
most data pairs in the PDF estimate.

\section{How to interpret the TDMI for a population, or, TDMI-based
  methods for interpreting population diversity}
\label{sec:tdmi_based_methods}

To achieve a practical understanding of the meaning of the TDMI in the
context of a population, we have to combine information from the
previous section to construct an explicitly numerically computable
means of interpreting $\bar{I}(\tau)$ and $\hat{I}(\tau)$. Practically speaking,
there are two broad situations: (i) $\bar{I}(\tau)$ is practically
calculable (when $\bar{I}(\tau)$ is calculable, $\hat{I}(\tau)$ always will be);
and (ii), $\bar{I}(\tau)$ is not calculable (usually to estimate $\bar{I}(\tau)$ there
need to be at least $100$ \emph{pairs of points} per representative
element) leaving us only with $\hat{I}$-related quantities. Relative
to the first situation, define the difference between $\bar{I}(\tau)$ and
$\hat{I}(\tau)$, or
\begin{align}
\label{equation:split_divergence_sources}
\delta I(\tau) &= | \bar{I}(\tau) - \hat{I} (\tau)| \\ \notag
&= |\int_{\mathcal{\bar{S}}} p_1(\mathcal{\bar{S}}) -
\int_{\hat{S}}p_1(\hat{S})| +  |\int_{\mathcal{\bar{S}}}\bar{G} - \int_{\hat{S}} \hat{G} |+
(\bar{B} - \hat{B}) | \\ \notag
&= \delta \rho + \delta G_{\int} + \delta B
\end{align}
This allow for the following conjecture which we will not proven in
this paper:
\begin{conjecture}
  In the circumstance where $\bar{I}(\tau)$ can be accurately
  estimated, $\delta I(\tau) \sim 0$ if and only if the population
  used to estimate $\bar{I}(\tau)$ and $\hat{I}(\tau)$ is
  statistically homogeneous temporally (i.e., the PDFs representing
  the individuals in the population are identical, as are the PDFs
  under temporal evolution).
\label{conjecture:1} 
\end{conjecture}
The forward direction of the if and only if statement, that $\delta I(\tau)
\neq 0$ implies a heterogeneous population will be
briefly discussed in section \ref{sec:graph_dependent_effects_on_I};
this direction is more complicated to prove. The reverse direction of the
if statement in this conjecture claims that if the population
represents a single, stationary, homogeneous distribution then $\delta
I(\tau) \sim 0$; this claim relies on the fact that in this circumstance all
$\epsilon$'s are zero and thus $\bar{I}(\tau)$ (Eqn. \ref{equation:ibar})
and $\hat{I}(\tau)$ (Eqn. \ref{equation:aggregated_MI_def}) represent a
homogeneous source and are equivalent up to bias. Essentially, when
one can estimate $\delta I(\tau)$, one can interpret the population make-up
\emph{without} delving deeply into the detailed \emph{sources} of the
TDMI. In contrast, when only $\hat{I}(\tau)$ is practically calculable, the
interpretation of $\hat{I}(\tau)$ can only be understood though
understanding the \emph{source} of the TDMI. Nevertheless, in general,
it is insightful to understand the sources of the TDMI, and the
sources of the TDMI are tied to the make-up of the population.

From a detailed perspective, the make-up of the population is
important because the deviation of the TDMI from the homogeneous case
is due to non-zero $\epsilon$'s, and the source of non-zero
$\bar{\epsilon}$'s can differ from the source of non-zero
$\hat{\epsilon}$'s. Specifically, $\bar{\epsilon}$ can \emph{only} be
non-zero because of differences between the graphs of the
$p_i$'s. This is because \emph{all the $p_i$s for the average TDMI
  have the same support.} In contrast, the source of non-zero
$\hat{\epsilon}$'s is due to a heterogeneous population can be split
into three broad categories: (i) differences in the TDMI estimates due
to differences in the supports \emph{independent} of the graphs of the
PDFs; (ii) differences in the TDMI estimates due to differences in the
graphs \emph{independent} of the supports; and (iii), differences in
the TDMI estimates due to the supports' \emph{effect on the graphs}.

\begin{figure}
    \subfigure[$p_i(S_i)$ for three distributions of Gaussian random
      numbers with means equal to $0$, $2$ and $4$]
    {
      \epsfig{file=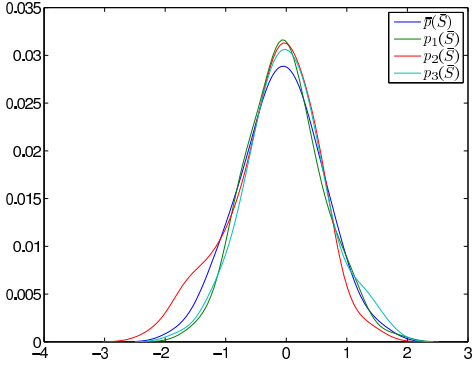, height=10cm}
      \label{figure:p_ave_normalized_support}
    }
    \subfigure[$p_i(S)$ for three distributions of Gaussian random
      numbers with means equal to $0$, $2$ and $4$ as well as
      $p(s)=1/3 \sum_{i=1}^3 p_i(\hat{S})$.]
    {
      \epsfig{file=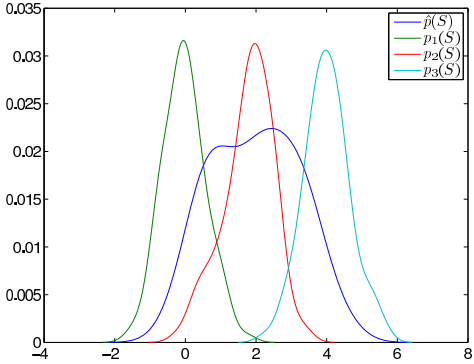, height=10cm}
      \label{figure:p_agg_full_support}
    }
    \caption{Graphically comparing $\bar{p}$ (average PDF) and
      $\hat{p}$ (PDF of the aggregate) for a collection of three
      collections of Gaussian random numbers whose distributions have
      means $0$, $2$ and $4$ respectively.}
    \label{figure:p_mean_agg_vs_ave}
\end{figure}

\subsection{Support dependent, graph independent,
  effects on the population TDMI}
\label{sec:support_effects_interp}

To understand and \emph{quantify} the differences in the TDMI
estimates due to differences in the supports \emph{independent} of the
graphs of the PDFs, consider the difference between the random
permutation bias estimates defined in section
\ref{sec:RP_bias_estimates}.  

First, recall that the population-wide random permutation bias
estimate will be roughly equivalent to the estimator bias, or $B_{PRP}(\tau)
\approx B_E(\tau)$ regardless of the supports or densities of the elements
(c.f., \cite{estimation_of_MI_EHR} for small sample size
qualifications of this statement). Next, note that the individual-wise
random permutation bias estimate, $\hat{B}_{IRP}(\tau)$ represents the
\emph{bias due to heterogeneity in the supports} plus the estimator
bias. Thus, the contribution to the bias due to the diversity in
population normalization is approximated by the difference between the
individual-wise and population-wise random permutation bias estimates:
\begin{equation}
B_{RP}(\tau) = |\hat{B}_{PRP}(\tau) - \hat{B}_{IRP}(\tau)|.
\end{equation}
There are two reasons why $B_{RP}(\tau)$ can be non-zero.  First the number
of points used to calculate the two can differ by orders of magnitude
(say, a population of $1,000$ with $10$ points each); in this case,
$B_{RP}(\tau)$ represents the $1/n$ effect on the bias estimates. In the
case where the number of pairs used to estimate $\hat{B}_{PRP}(\tau)$ and
$\hat{B}_{IRP}(\tau)$ are relatively similar (e.g., more than $100$ and within an
order of magnitude; to control for the number of points, it is easy
reduce the cardinality of the set used to calculate $\hat{B}_{PRP}(\tau)$)
Fig. \ref{figure:p_agg_full_support} shows visually how these bias
estimates would render differently. In this context, $\hat{B}_{IRP}(\tau)$ would
be identical to $I$, where as randomly permuting the entire
population, such as is done to estimate $\hat{B}_{PRP}(\tau)$, will result in one
of the marginal distributions becoming $\hat{p}(\hat{S})$ --- a
uniform distribution instead of three Gaussians with distinct means
--- thus greatly changing the amount of mutual information. These
effects are primarily support-driven effects; note that while it is
possible that differences in the underlying distribution function can
be rendered through $B_{RP}(\tau)$, differences in the support of those
distributions will always be rendered through $B_{RP}(\tau)$. As we will see
in a moment, $B_{RP}(\tau) \approx \hat{B}_{IRP}(\tau)$ is \emph{not enough} to imply
that $\delta I(\tau) \approx 0$, but is enough to imply that the variance in
the boundaries of the supports will all be relatively
small. Nevertheless, while in some circumstances it may be difficult
to use the bias estimates to detect a difference in the average versus
aggregate TDMI, we can use the bias estimates to interpret the average
and aggregate TDMI signal. In particular, when $B_{RP}(\tau) \leq B_E(\tau)$,
intermixing individuals' measurements has no effect on the random
permutation bias estimate, implying that there is very little
population selection information in the TDMI estimate. Thus, $B_{RP}(\tau)
\leq B_E(\tau)$ \emph{at least} implies overlapping distribution supports.
Similarly, when $B_{RP}(\tau) \gg B_E(\tau)$, intermixing elements has a profound
effect on the random permutation bias estimates; in this instance,
$B_{RP}(\tau)$ reveals a bias whose source is the diversity of the supports
among the elements. This leads us to the measure of homogeneity of
supports that is very computable even for poorly measured populations
(e.g., when only $\hat{I}(\tau)$ is calculable); the \emph{TDMI homogeneity
  of support} is defined by the following equation:
\begin{equation}
\mathcal{H}_{S}(\tau)  =\frac{|\hat{B}_{IRP}(\tau) - \hat{I}(\tau)|}{\hat{I}(\tau)}
\end{equation}
The closer $\mathcal{H}_{S}(\tau)$ is to one, the less the diversity of the
supports over the population; similarly, the closer $\mathcal{H}_{S}(\tau)$
is to zero, the greater the diversity of the supports over the
population. (Again, note one must control for the dependence on the
number of pairs used to estimate the above quantities.)

It is worth noting that a similar analysis can by done by comparing
$\hat{\mathcal{I}}(\tau)$ to $\hat{I}(\tau)$, as their difference will reveal
support based effects. The principles behind a $\delta \hat{I}(\tau) =
|\hat{\mathcal{I}}(\tau)-\hat{I}(\tau)|$ and $\mathcal{H}_S(\tau)$ are similar in that
they both address normalization of support based effects, only
$\mathcal{H}_S(\tau)$ depends on quantities that represent distributions ---
$\hat{B}_{IRP}(\tau)$ and $\hat{B}_{PRP}(\tau)$ can both be estimated many times
--- and thus are likely more robust.

\subsection{Graph dependent, support independent,
  effects on the population TDMI}
\label{sec:graph_dependent_effects_on_I}

To understand in detail how differences in the graphs
\emph{independent} of the supports can affect the $\bar{I}(\tau)$ and $\hat{I}(\tau)$,
begin by assuming that all the $p_i$'s \emph{have the same support}, or that
$\cap_{i=1}^N S_i = \cup_{i=1}^N S_i$. In this circumstance, the
$\bar{\epsilon}_i = \hat{\epsilon}_i$ for all $i$. Thus, the
contribution of the diversity of PDFs within the population to $I$, or
the deviation from the mean at a particular $x \in S$ value, is
captured by $\bar{G}(\tau)$ and $\hat{G}(\tau)$ as defined in
Eqs. \ref{equation:gbar} and \ref{equation:ghat}. Consequently, the
only way that $\bar{I}(\tau)$ can be different from $\hat{I}(\tau)$ up to the
estimator bias is for the variation in the collections of PDFs to be
due to the \emph{order of averaging} as rendered though the $G$'s.

Based on the aforementioned intuition, we claim (e.g., conjecture
\ref{conjecture:1}) that $\delta I(\tau)$ is equal to zero if and only
if all the $\epsilon$'s are zero. While we will not present a
qualified proof of this claim here, we can offer an intuitive argument
as to why our claim is justified. First note that by inspection, if
$\epsilon_i=0$ for all $i$, $\delta G(\tau) = \bar{G}(\tau) =
\hat{G}(\tau) = 0$. Now, what remains is to understand what happens to
the $G$'s when there are non-zero $\epsilon$'s; to do this, note that
we reduce the $G$'s to the terms they do not have in common:
\begin{align}
\label{equation:simple_gs}
&\bar{G}(\tau) \sim \bar{g}(\tau) = (p_{i \tau} - [\epsilon])
\log (  \frac{1-\frac{[\epsilon]}{p_{i
      \tau}}}{(1-\frac{[\epsilon]}{p_{i}})(1-\frac{[\epsilon]}{p_{\tau}})}
) \\
&\hat{G}(\tau) \sim \hat{g}(\tau) = \frac{1}{N p_{i \tau}} \sum_{j=1}^N
(p_{i \tau} - \epsilon_j) \log (  \frac{1-\frac{\epsilon_j}{p_{i
      \tau}}}{(1-\frac{\epsilon_j}{p_{i}})(1-\frac{\epsilon_j}{p_{\tau}})}
)
\end{align}
and then consider the difference in these quantities:
\begin{equation}
\delta G \sim \delta g(\tau) = |\bar{g}(\tau) - \hat{g}(\tau)|.
\end{equation}
Now, further noting that $\bar{g}(\tau)$ is convex (or concave,
depending on the $p$'s) and applying standard convexity arguments,
$\delta{g}$ will not equal zero unless $\epsilon_i =0$ for all
$i$. Thus, while it is possible that, through the act of integrating
the $G$'s, symmetries will allow for the $G$'s to be equal, it is
extremely unlikely. Moreover, because the convexity or concavity of
$\bar{g}(\tau)$ depends on the nature of the $p$'s, it is difficult to
say whether $\bar{g}(\tau)$ will be, in general, greater or less than
$\hat{g}(\tau)$. Nevertheless, it appears in computational experiments
that $\hat{g}(\tau)$ is often less than $\bar{g}(\tau)$. In any event,
it is now more clear how diversity amongst the distribution of $p$'s
over the same support can (and likely will) force $\delta I(\tau) \neq
0$.

In the situation where $\bar{I}(\tau)$ is not accessible, it may not be
possible to fully understand the meaning of $\hat{I}(\tau)$.  While
$\mathcal{H}_S(\tau)$ can help identify support based effects, pure
graph-based \emph{temporally dependent} effects may be difficult to
estimate. In particular, if the sample size for some of the
individuals is small, then it will be difficult to determine the
contribution to $\hat{I}(\tau)$ due to purely graphic diversity simply
because there will be such high variance in the graphical PDF
estimates due to small sample sizes\footnote{To see the variation in
  the PDF estimates due to small sample sizes, observe the PDF
  estimates for different sets of uniform random numbers with small
  cardinality.}. In this case, the best that can be done is to
estimate more static measures of graphic diversity such as those
presented in section \ref{sec:non_TDMI_hetero_methods}.

\subsection{Support dependent, graph-based
effects on the population TDMI}

There are two potential contributors to support dependent, graph-based
effects on $\delta I(\tau)$, $\delta G(\tau)$ and $\delta \rho(\tau)$.

The contribution to $\delta I(\tau)$ due to $\delta \rho(\tau)$ is entirely due to
the limits of integration; the integrand for the average and aggregate
$\rho$ component of the TDMI are identical. Thus, intuitively, $\delta
\rho>0$ because of the \emph{relative location} of the support of
$p_1$ in reference to the total support of the population; $p_1$ will
represent a \emph{more peaked} distribution when defined on $\hat{S}$
compared to $\mathcal{\bar{S}}$. Note that while $\delta \rho$ is, in
general, computable, it has similar characteristics to
$\mathcal{H}_S(\tau)$ with more severe bias issues.

The contribution due to $\delta G_{\int}$ is not as intuitive; to
understand how diversity in the supports contributes to $\delta
G_{\int}$ via the induced differences in the $\epsilon$'s, consider
Figs. \ref{figure:p_ave_normalized_support} and
\ref{figure:p_agg_full_support}. Relative to
Fig. \ref{figure:p_ave_normalized_support}, begin by defining
$\bar{p}(\mathcal{\bar{S}})$ as the average of the PDFs relative to
the abstract support, or
$\bar{p}(\mathcal{\bar{S}})=\frac{1}{3}(p_1(\mathcal{\bar{S}}) +
p_2(\mathcal{\bar{S}}) + p_3(\mathcal{\bar{S}}))$; here all the
$\bar{\epsilon}_i$'s will be small and independent of the
support. This is how variation in the population is rendered when
calculating $\bar{I}$, and thus how $\bar{G}$ will render. In
contrast, define the average of the PDFs relative to the \emph{total
  support}, or $\hat{p}(\hat{S}) =\frac{1}{3}(p_1(\hat{S})
+p_2(\hat{S}) + p_3(\hat{S}))$; this is the aggregate scenario. Here
it is clear that \emph{both} the averaged PDF will not resemble
\emph{any} of the PDFs \emph{and} relative to a selected
$p_1$. Moreover, all $\hat{\epsilon}_i$'s will be relatively large and
on the order of the various $p_i(\hat{S})$'s over a non-trivial
portion of the population support $\cup_{i=1}^N S_i$. Because of this,
when the supports of the individuals differ, the largest term in
$\hat{I}(\tau)$, $\hat{G}(\tau)$, will be accounting primarily for
\emph{variation within the distribution of the supports of the
  population}, rather than support-independent variation within the
population. Moreover, when the supports of the individuals are
relatively invariant, $\hat{I}$ will be independent of time even when
the $I$ of an individual varies with $\tau$. In any event, the point
is, variation in the supports of otherwise identical distributions
affects how the distributions are rendered though the TDMI
calculation.

Finally, when only $\hat{I}(\tau)$ is available, which implies the
presence of individuals with too few pairs of points to accurately
estimate a PDF and thus the TDMI, and when there are support-dependent
graph-based effects in the TDMI, it will likely be difficult to
separate the support dependent, graph-based effects from the support
\emph{independent} graph-based effects on the TDMI (e.g., on the
structure of the temporal correlation).

\section{Non-TDMI-based methods for interpreting population diversity}
\label{sec:non_TDMI_hetero_methods}

In this paper, we claim that the TDMI-based analysis can be used to
both detail nonlinear correlation in time \emph{and} interpret the
composition of the population to which that correlation pertains to
(i.e., whether the TDMI reflects and individual/homogeneous population
or a diverse population). To verify this claim, we require a set of
methods for establishing a \emph{baseline} that are independent of
information-theoretic machinery and can be used to interpret the
make-up of the population. We propose three different quantifications
of homogeneity of a population: (i) homogeneity in measurement
representation, which addresses the variance in the distribution of
the number of measurements per element of the population; (ii)
homogeneity in support, which addresses variation in the supports of
each elements' distribution; and (iii) homogeneity in density, which
addresses variation in the PDFs (or the graphs of the PDFs) over the
population. \emph{Note} that all but one of the methods for
quantifying homogeneity are \emph{independent of time}, and all are
\emph{independent of any time-based correlation structure} existent
within the data set. Moreover, the homogeneity qualification methods
we propose here are neither exhaustive nor particularly innovative;
rather they are simple intuitive methods devised to interpret and
confirm the TDMI-based results. Nevertheless, many of these methods
are useful in their own right; moreover, at least one of the
quantities we define here is required to supplement the TDMI analysis
when very few measurements exist per individual. Finally, table
\ref{table:tdmi_heuristic_metrics_summary} contains a summary of the
ten TDMI-independent quantities are we use to verify the TDMI
methodology.

\renewcommand{\arraystretch}{1.5}
\begin{table}
\begin{tabular}{|l||p{3cm}|p{4cm}|}
  \hline
  \multicolumn{3}{|c|}{non-TDMI-based quantities for characterizing a
    population} \\
  \Cline{2pt}{1-3}
  $H_{\bar{x}}$ & difference between the population and individual
  element means & $\sim 0$ implies \emph{either} (i) most
  elements have a similar number of measurements, or (ii) the
  individuals come from distributions with similar means; $\gg 0$
  implies the converse \\
  $V(f(n))$ & variance of the PDF of the number of measurements per individual &  (i) $V\sim 0$, $H_{\bar{x}} \sim 0$ imply elements
  were measured similarly; $ \gg 0$, $H_{\bar{x}} \sim
  0$ implies elements measured at different rates; $\gg 0$, $H_{\bar{x}}
  \gg 0$ implies elements measured at different rates with
  differing source distributions. \\
  $\bar{s}_{min}$ & $ E[s_{min}(i)]$  & lower support boundary mean. \\
  $\bar{s}_{max}$ & $E[s_{max}(i)]$  & upper support boundary mean. \\
  $V_{s_{min}}$ & $Var(s_{min})$ & lower support boundary variance. \\
  $V_{s_{max}}$ & $Var(s_{max})$ & upper support boundary variance.  \\
  $\bar{|S|}$ & $\bar{s}_{max} - \bar{s}_{min}$ & length of support
  mean. \\
  $V_{\bar{|S|}}$ & $Var(\bar{s}_{max} - \bar{s}_{min})$ & length of
  support variance.\\
  $H_{RA}$  & area between the (point-wise) least and greatest PDF
  graph & quantifies variance between the PDFs of the population; $\sim
  0$ implies element PDFs are homogeneous; very
  sensitive. \\  
  $V_S(p)$ & $\int_S E[(p(x))^2] - E[p(x)]^2 dx$, variance of the
  PDFs \emph{relative to a specified support}, $S$ & $\sim 0$ implies homogeneity in PDFs; larger $Var_S(f)$ implies greater heterogeneity in the PDFs.\\  
  $V_{\hat{S}}(p)$ & $V_S(p)$ calculated relative to the support of
  the aggregate population; $\hat{S} = \cup_{i=1}^N \hat{S}_i$; note
  that there does exist an aggregate normalized support,
  $\mathcal{\hat{S}}$, but we will not use this quantity here. & $V_{\hat{S}}(p)$  has
    the same interpretation as $V_S(p)$ in general, but has the
    potiental to \emph{include}
    support-based effects. \\  
  $V_{\bar{\mathcal{S}}}(p)$ & $V_S(p)$ calculated relative to the abstract support of
  the population, $\bar{\mathcal{S}}$ & $V_{\bar{\mathcal{S}}}(p)$ has
    the same interpretation as $V_S(p)$ in general, but excludes
    support-based effects.  \\   
 \hline
\end{tabular}
\caption{Summary of all the non-TDMI based metrics used to assess
  homogeneity in a population (both among the graphs and the supports)
used to verify the TDMI-type analysis.}
\label{table:tdmi_heuristic_metrics_summary}
\end{table}

\subsection{Homogeneity in measurement composition}

To quantify \emph{homogeneity in measurement composition}, begin with
the following two quantities.  First, consider the difference between
the mean of the raw measurements over the population versus the mean
of the individual-wise measurement means, or:
\begin{equation}
\begin{split}
H_{\bar{x}} = &\left( \frac{1}{\sum_{k=1}^N n_k} \sum_{i=1}^{\sum_{k=1}^N n_k} x_i \right) \\
&- \left( \frac{1}{N}
\sum_{k=1}^{N} \frac{1}{n_k} \sum_{i=1}^{n_k} x_{i + \sum_{j=0}^{k-1} n_j } \right) 
\end{split}
\end{equation}
where $n_k$ is the number of points contributed by individual $k$, $N$
is the number of individuals in the population, and $n_0 =0$. Now,
$H_{\bar{x}} \approx 0$ under two circumstances: (i) the distribution
of $n_k$'s has zero or small variance, \emph{regardless} of the
collection of individual distributions; or (ii) each individual comes
from an identical distribution.  Second, consider the variance of the
probability density of the number of measurements per individual:
\begin{equation}
V_{f(n)} = \text{Var}(f(n))
\end{equation}
where $f(n)$ denotes the density of measurements per individual.
Combining these two quantities we arrive at three cases: (i) $V_{f(n)}
\sim 0$ implies that $H_{\bar{x}} \sim 0$, together implying that the
elements were measured similarly --- no insight into the original
distributions can be made; (ii) $V_{f(n)} \gg 0$ and $H_{\bar{x}} \sim
0$ together imply that the elements were measured at different rates
regardless of their source distributions (which can be identical); and
(iii) $V_{f(n)} \gg 0$ and $H_{\bar{x}} \gg 0$ together implies that
the elements were measured at different rates and likely have
differing source distributions. Note, that in general, both of these
metrics are rather sensitive to diversity in a population.

\subsection{Homogeneity in measurement distribution supports}

To characterize \emph{homogeneity in distribution support} we rely on
a brute force homogeneity characterization technique. Begin by
recalling that the support for element $i$'s distribution as $S_i =
[s_{min}(i), s_{max}(i)]$. Given these sets, which are defined by the
individuals' measurements, define the mean and variance of the support
minima, maxima, and length by:
\begin{align}
\bar{s}_{min} &= E[s_{min}(i)] \\
\bar{s}_{max} &= E[s_{max}(i)] \\
V_{s_{min}} &= Var(s_{min}) \\
V_{s_{max}} &= Var(s_{max}) \\
\bar{|S|} &= \bar{s}_{max} - \bar{s}_{min} \\
V_{\bar{|S|}} &= Var(\bar{s}_{max} - \bar{s}_{min})
\end{align}
These quantities afford relatively simple representations. For
instance, when the minima, maxima, and lengths for the population have
small variance, the intersection of the supports will not differ
significantly from the union of the support --- meaning the supports
overlap.  While a large variance in any either the minima, maxima, or
lengths implies that the supports differ significantly over the
population.

\subsection{Homogeneity in the distribution of the graphs of the
  measurement PDFs}

To specify \emph{homogeneity in the PDF} of the population we will use
two methods.  Intuitively, all of the methods characterize, in one way
or another, the \emph{width} of the maximum and minimum band of PDFs
of the population over the support of the entire population.  Begin by
defining the PDF for an individual by $p_i(x)$, the supremum of the PDFs
of the population by $\max_i (p(x)) = p_M(x)$, and the infimum of PDFs
of the population by $\min_i (p(x)) = p_m(x)$, over the union of the
supports, $S = \cup_i^N S_i$. First, using the $L_1$ (pseudo)
distance \footnote{Note, the $L_1$ difference is not technically a
  distance function or a metric because it does not satisfy the
  triangle inequality.} we can define the \emph{relative area} of the
width of the band of PDFs by:
\begin{equation}
H_{RA} = \frac{\int_S |p_M(x) - p_m(x)| dx}{\int_S p_M(x) dx}
\end{equation}
The relative area, $H_{RA}$ is literally the proportion of the
supremum of the collection of PDFs that coincides with the infimum of
the collection of PDFs.  When $H_{RA}$ is close to one, the maximum
distance between PDFs over the population occupies all the volume of
the population-wide PDF.  In other words, the population has at least
two substantially different PDFs. Similarly, when $H_{RA}$ is near
zero, this implies that the proportion of the area between the
supremum and infimum over the collection of $p_i$'s relative to the
total area occupied by the supremum of the $p_i$'s over the population
is very small.  Thus the \emph{implication} of $H_{RA}$ being near
zero is that the $p_i$'s are all nearly identical. \emph{However},
this method is very sensitive to heterogeneity; \emph{a single}
individual's PDF differing from the rest of the population can
maximize $H_{RA}$ at one. In contrast, the second method for
evaluating the diversity in PDFs over the population quantifies
diversity from a mean within the population by estimating the
\emph{variance of the PDFs at a given at a given $x$} integrated over
a given support ($S$), or
\begin{equation}
V_S(p) = \int_S E[(p(x))^2] - E[p(x)]^2 dx
\end{equation}
\emph{Note}, $V_S(p)$ can be estimated relative to \emph{two
  different} supports, the union of the supports, or the
abstract support. This is an $L_2$ flavored
representation of the variation in PDFs; the variance of the $p_i$'s
at a given $x$ is maximized when $p_i$'s are maximally orthogonal (in
the sense of an inner product between the $p_i$'s) to one another, and
minimized when the $p_i$'s are minimally orthogonal (meaning they
coincide). Thus, $V_S(p)$ has the potential to capture both support-
and graph-based variation, depending on whether $V$ is calculated
relative to $\hat{S}$, which will include support-based effects, or
$\bar{\mathcal{S}}$, which will not include support-based effects.

\section{Assembling the pieces: an explicit prescription for TDMI analysis and interpretation for a
population of time series for a fixed time separation $\delta t$}
\label{sec:interpretation_summary}

\renewcommand{\arraystretch}{1.5}
\begin{table}
\begin{tabular}{|l||p{3cm}|p{4cm}|}
  \hline
  \multicolumn{3}{|c|}{TDMI-based analysis quantities} \\
  \Cline{2pt}{1-3}
  Quantity & What it signifies & What it quantifies \\
  $\bar{I}(\delta t)$ & population averaged TDMI & quantifies average TDMI of a population \\
  $\hat{I}(\delta t)$ & aggregated population TDMI & quantifies TDMI of an
  aggregated population \\
  $\hat{\mathcal{I}}(\delta t)$ & aggregated population calculated relative to the
  abstract support $\hat{\mathcal{S}}$ & support independent TDMI of an
  aggregated population \\
  $\delta I(\delta t)$ & $|\hat{I}(\delta t)-\bar{I}(\delta t)|$; difference between the average and
  aggregate TDMI & $\sim 0$ implies homogeneity, $<0$
  implies heterogeneity \\
  $B_E(\delta t)$ & PDF estimator bias; usually $B_E(\delta t) \sim
  B_{PRP}(\delta t)$; $B_E(\delta t)$ can be
  estimated in a variety of ways & the number above which the $I$
  is considered to be positive \\
  $\bar{B}_{IRP}(\delta t)$ & individual permutation bias averaged over a population & bias estimate that preserves
  information about the relative ranges of individuals \\
  $\hat{B}_{IRP}(\delta t)$ & individual permutation bias & bias estimate that preserves
  information about the relative ranges of individuals \\
  $\hat{B}_{PRP}(\delta t)$ & population permutation bias & bias estimate that destroys
  information about the relative ranges of individuals \\
  $\mathcal{H}_S(\delta t)$ & $\frac{|\hat{B}_{IRP}(\delta t)-\hat{I}(\delta
    t)|}{\hat{I}(\delta t)}$;
  quantifies diversity of supports &
  $\sim 1$ implies homogeneous supports; $\sim 0$ implies diverse supports \\
  $B_{RP}(\delta t)$ &$|\hat{B}_{PRP}(\delta t)-\hat{B}_{IRP}(\delta t)|$; quantifies diversity of supports;
  quantifies cardinality of individual data sets &
  $\sim \hat{B}_{IRP}(\delta t)$ can imply diverse supports or cardinality per-element
  data sets; $\sim 0$ can imply homogeneity in supports \\
  $\delta G(\delta t)$ & \emph{difference in the difference} between how
  population diversity renders in $\bar{I}$ and
  $\hat{I}(\delta t)$ & $>0$ implies population diversity \\
  $\delta \rho(\delta t)$ &  $|\int_{\mathcal{\bar{S}}} p_1(\mathcal{\bar{S}}) -
  \int_{\hat{S}}p_1(\hat{S})|$; quantifies diversity in supports &  $>0$
  implies population diversity. \\
  $H_{\Theta}(\delta t)$ & how representative the population used to estimate $I$
  at $\delta t$ is of the time-independent (e.g., the entire) population
  & $\sim 0$ implies the entire population is well represented; $\sim 1$
  implies portions of the population are overrepresented \\
  $N_{min}(\delta t)$ & minimum number of \emph{pairs} of points
  contributed by any one individual & a lower bound on the representation of an
  individual; $1/N_{min}(\delta t)$ is a rough
  estimate of $B_E(\delta t)$ for the individual with the fewest pairs \\
  \hline
\end{tabular}
\caption{Summary of all the TDMI-based metrics used to interpret the
  TDMI and determine the population composition.}
\label{table:tdmi_metrics_summary}
\end{table}

\begin{figure*}[tbp]
  \epsfig{file=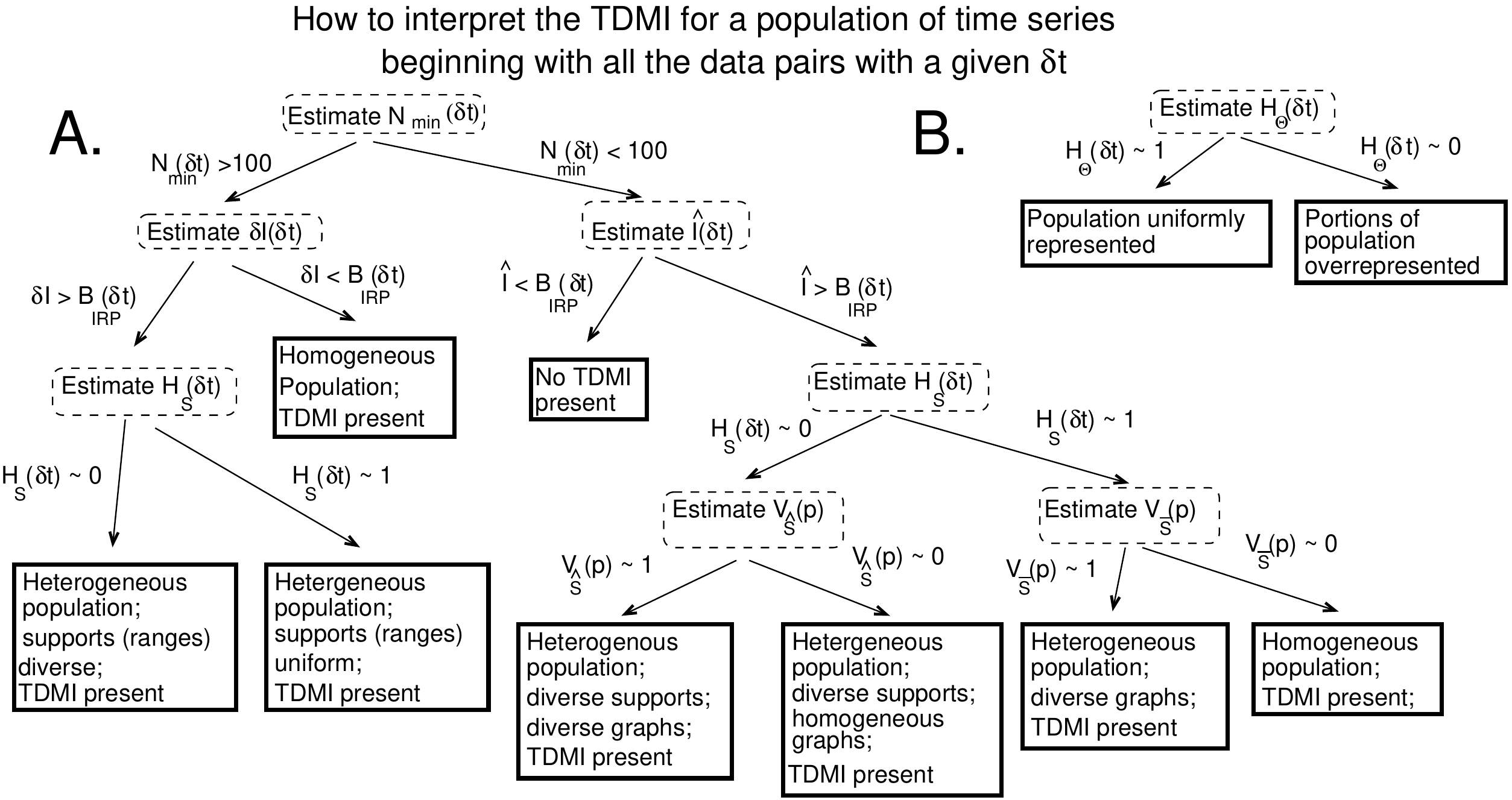, height=10cm} 
\caption{The graphical schematic for the TDMI analysis of a population.}
\label{fig:TDMI_analysis_schematic}
\end{figure*}

The interpretation of the TDMI and entropy for a complex, diversely
measured population can be split into three broad steps: (i)
performing a preliminary interpretation of $\bar{I}(\delta t)$ and
$\hat{I}(\delta t)$; (ii) performing an interpretation of $\delta
I(\delta t)$ or $\hat{I}(\delta t)$ for the population; and (iii)
understanding the make-up of the data explicitly used to estimate the
PDFs, yielding an understanding of what proportion of the population
as used in the calculation. All the TDMI quantities used for the
TDMI-based analysis are shown in table
\ref{table:tdmi_metrics_summary}, a graphical schematic for applying
this infrastructure is shown in
Fig. \ref{fig:TDMI_analysis_schematic}, and a detailed algorithmic
schematic for applying the TDMI infrastructure to a population is
depicted via pseudocode in appendix \ref{sec:app_pc}.

\subsection{Step one: determining the computability of $\bar{I}(\delta t)$}

To begin, one must determine whether $\bar{I}(\delta t)$ and
$\hat{I}(\delta t)$ are calculable for a given (or set of) $\delta
t$(s). In general, to estimate $\bar{I}(\delta t)$ \emph{every
  representative individual} must (under most circumstances) have at
least \emph{$100$ pairs of points} available for the TDMI calculation
\cite{estimation_of_MI_EHR}. Similarly, to estimate $\hat{I}(\delta
t)$ there must be at least $100$ pairs of points \emph{gathered over
  the entire population}---this is why $\hat{I}(\delta t)$ is so
useful in the context of a population.

Assuming that $\bar{I}(\delta t)$ is calculable, because the
calculation of $I$ for an individual \emph{is independent of the
  support of the distribution}, the variance in the distribution of
$\bar{I}(\delta t)$ is due to differences in the \emph{graphs} of the
PDFs representing each patient at a given $\delta t$. Further, because
$\bar{I}(\delta t)$ is made of individuals who have been averaged, the
interpretation of the statistical moments of $\bar{I}(\delta t)$
(i.e., the mean, variance, etc), is a scientific problem that depends
on the particular circumstances.

The interpretation of $\hat{I}(\delta t)$ is more difficult because
$\hat{I}(\delta t)$ can be composed of purely graphical, purely
support, and intermixed support and graphical components, Thus,
because $\hat{I}(\delta t)$ is a population-dependent quantity where
the individual contributions cannot be separated, it will be treated
in the next section with $\delta I(\delta t)$.

\subsection{Step two (A in Fig. \ref{fig:TDMI_analysis_schematic}):
  interpreting $\delta I(\delta t)$ or $\hat{I}(\delta t)$}

Step two has two courses of action depending on whether it is possible
to calculate $\bar{I}(\delta t)$ or not: (i) $\bar{I}(\delta t)$ and
$\hat{I}(\delta t)$ are calculable and thus $\delta I(\delta t)$ can
be computed; and (ii) only $\hat{I}(\delta t)$, $B_{RP}(\delta t)$,
and $\mathcal{H}_S(\delta t)$ are calculable (when $\hat{I}(\delta t)$
is calculable, this will always be the case). When $\delta I(\delta
t)$ is available, it, as estimated by both a KDE and histogram
estimator, is all we need know: the closer $\delta I(\delta t)$ is to
zero, the more homogeneous the population is and the more
$\hat{I}(\delta t)$ represents a single, statistically singular source
and the larger in magnitude $\delta I(\delta t)$ is, the more
statistically heterogeneous the population is and the more
$\hat{I}(\delta t)$ represents the population.  Of course, if the
histogram and KDE TDMI estimates differ substantially, it is likely
that there are significant small sample size effects present in
$\bar{I}(\delta t)$, and this needs to be taken into consideration
when interpreting $\delta I(\delta t)$, $\bar{I}(\delta t)$ and
$\hat{I}(\delta t)$. Moreover, in this circumstance, calculation of
either $B_{RP}(\delta t)= |B_{IRP}(\delta t) - B_{PRP}(\delta t)|$ or
$\mathcal{H}_S(\delta t)$ can be used to further qualify the small
sample size effects on the variation in the supports versus the
graphs. Finally, when $\delta I(\delta t)$ is positive, and
$\mathcal{H}_S(\delta t)$ shows no diversity due to the supports, then
all the diversity in the population is due to the graph-based
diversity.

When $\bar{I}(\delta t)$ is not calculable, one is left with only
$\hat{I}(\delta t)$, $\hat{\mathcal{I}}(\delta t)$, and $B_{RP}(\delta
t)$ or $\mathcal{H}(\delta t)$. In this case, one can still use
$B_{RP}(\delta t)$ or $\mathcal{H}(\delta t)$ to detect the homo- or
heterogeneity in the supports. If there is no support-based variation
then pure graph-based heterogeneity maybe be difficult to determine;
in this circumstance we recommend using a non-TDMI metric such as
$V_S(p)$, which will have greater statistical power while sacrificing
temporal dependence, to help determine the graphical composition of
the population. In general, if there is support-based variation, it
will likely be difficult to separate support-based, versus
graph-based, contributions; it will be even more difficult to specify
the \emph{proportion} of diversity contributed by the support- versus
graph-based effects.

\subsection{Step three (B in Fig. \ref{fig:TDMI_analysis_schematic}): Assessing population representation}

Finally, it is extremely important to understand what portions of the
population \emph{actually} have points in a given $\delta t$ bin.
Recall that the make-up of the population used to estimate $I$ at a
specific $\delta t$ is a concern because of the filtering effect
(c.f., section \ref{sec:filter_as_bias}); specifically, it is possible
to have entire portions of the population excluded from the data set
as well has a highly nonuniform distribution of the population
represented in the data set used to estimate the PDFs. Written
differently, it is important to always remember that $\delta I$ is
always calculated relative to a \emph{fixed} $\delta t$ which will
have a particular bin population --- when studying the evolution of
$I$ as $\delta t$ is varied, \emph{the representative population can
  change as $\delta t$ changes}. Thus, it is important to at least
calculate $H_{\Theta}(\delta t)$ or an $H_{\Theta}$-like quantity to verify what
proportion of the population is being included in the PDF
estimate. Moreover, we also find it convenient to keep track of the
minimum (and sometimes maximum) number of pairs of points contributed
by an element represented in the data set used to estimate the PDFs;
we denote this number by $N_{min}(\delta t)$ as a measure of the
\emph{least} representative individual.

\section{Quantitative examples for TDMI interpretation and
  population homogeneity evaluation}
\label{sec:quantitative_examples}


\subsection{Simulated data examples: the quadratic map and the Gauss map}
\label{sec:quant_simulated_data} 

\begin{figure}[tbp]
\begin{center}
\subfigure[The graphs of the quadratic map
  (Eqn. \ref{equation:quadratic}) and the Gauss map
  (Eqn. \ref{equation:gauss})]
    {
      \epsfig{file=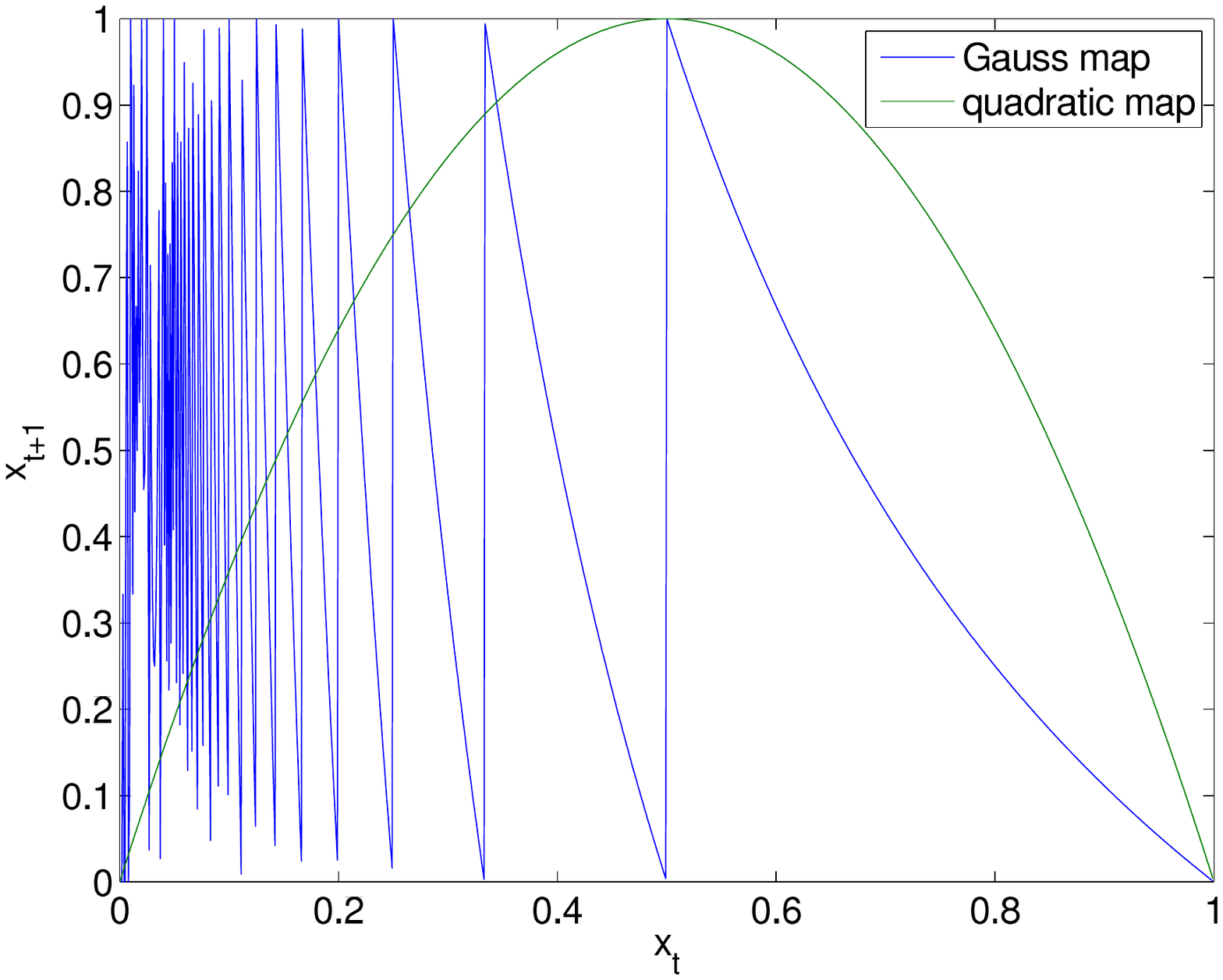, height=10cm}
      \label{figure:f_diff_agg_a}
    }
\subfigure[KDE of the invariant density (PDF of
the orbit) for the quadratic map, Gauss map, and the sum of the
quadratic and Gauss maps]
    {
      \epsfig{file=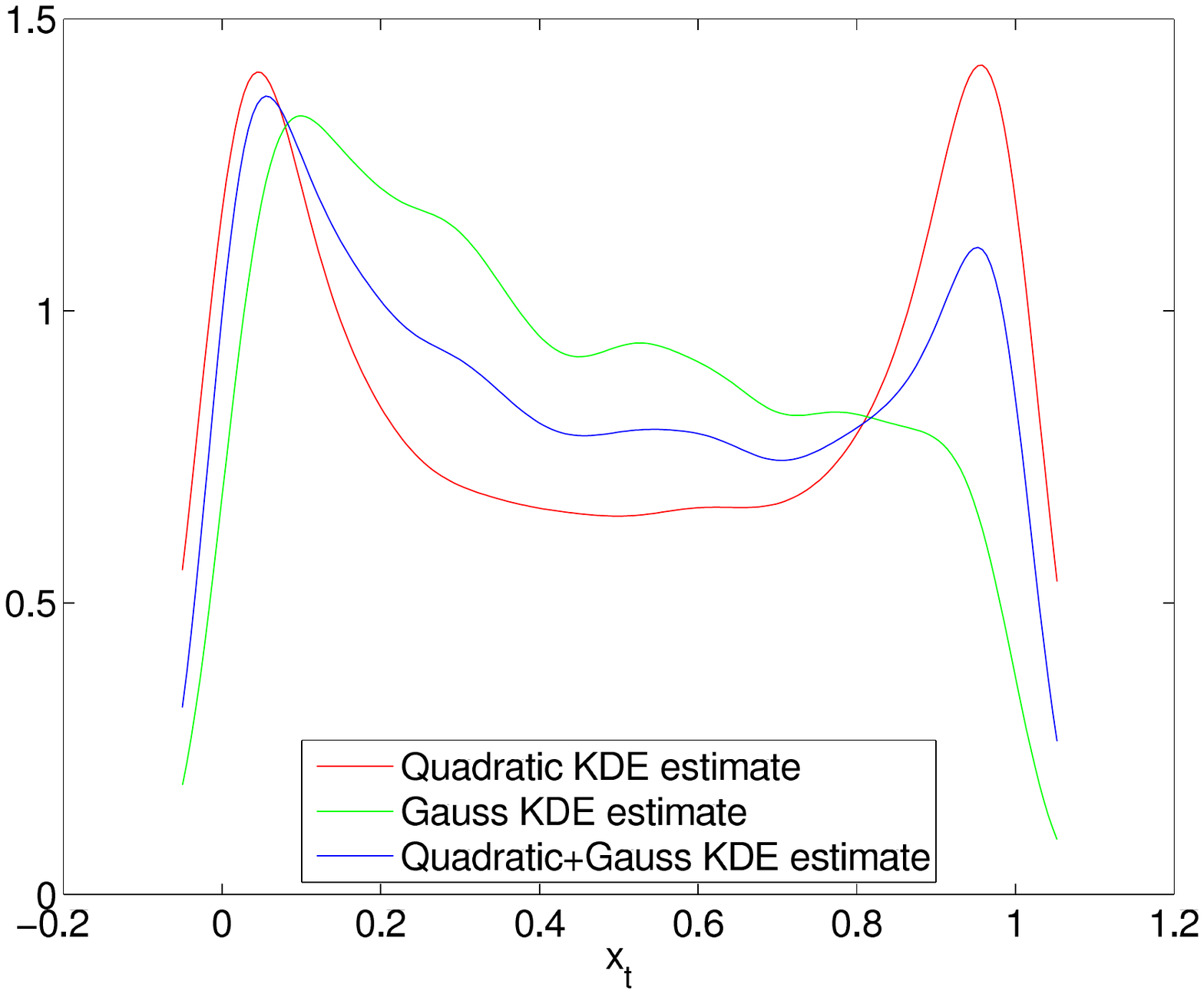, height=10cm} 
      \label{figure:f_diff_b}
    }
\end{center}
\caption{The graphs of the quadratic map
  (Eqn. \ref{equation:quadratic}) and the Gauss map
  (Eqn. \ref{equation:gauss}) --- note the significant difference
  between the graphs of the mappings, and invariant density (PDF of
  the orbit) for the quadratic map, Gauss map, and the sum of the
  quadratic and Gauss maps --- note the significant differences
  between the relative $p$'s.}
\label{fig:f_diff_agg}
\end{figure}

To explicitly demonstrate how to interpret $\bar{I}$ and $\hat{I}$ in
the presence of a diverse population in a variety of circumstances,
consider two sources of simulated data, the quadratic map
\begin{equation}
\label{equation:quadratic}
x_{t+1} = f(x_t) = ax_t(1-x_t)
\end{equation}
where $a$ is set to $4$ and the Gauss map
\begin{equation}
\label{equation:gauss}
x_{t+1} = g(x_t) = \frac{1}{x_t} \mod 1
\end{equation}
These sources were chosen because their statistical structures are
well understood \cite{graczyk_paper} \cite{jakobson_abscontin}
\cite{sprott_book}, they are chaotic, they are both $1$-dimensional
maps defined over the unit interval (meaning, they have the same
support), and they have relatively different invariant densities.
Figure \ref{fig:f_diff_agg} shows the the graphs of the quadratic and
Gauss maps, their individual invariant densities (PDFs of the orbit),
and the sum of their invariant densities.  Thus, in this context, the
difference between $p_f$ and $p_g$, $\epsilon(x)$, is both large
enough such that the $G$'s will be non-zero and is non-uniform over the
domain or nonlinearly dependent on $x$. The data sets we will use,
based on the maps above, include:
\begin{dataset}
 Quadratic map time-series with $1000$
  points; this is one of the data sets meant as a baseline from which all the
  other cases can be compared.
\end{dataset}
\begin{dataset}
 Gauss map time-series with $1000$
  points; this is one of the data sets meant as a baseline from which all the
  other cases can be compared.
\end{dataset}
\begin{dataset}
 Data sets $1$ and $2$ concatenated into a single
  data set with $2000$ data points; this data set is used primarily to
  test the effects of differing PDFs within a population on $\iota$,
  $G$, and thus, $\bar{I}$ versus $\hat{I}$.
\end{dataset}
\begin{dataset}
$50$ independent, concatenated quadratic map
  time-series with $20$ points each totally $1000$ points; this data
  set is meant to highlight the effect of the estimator bias when
  calculating $\bar{I}$ versus $\hat{I}$.
\end{dataset}
\begin{dataset}
$10$ independent, concatenated quadratic map
  time-series with $100$ points each totaling $1000$ points; this data
  set is meant to form a baseline for data set $6$.
\end{dataset}
\begin{dataset}
$10$ independent, concatenated quadratic map
  time-series with $100$ points with \emph{disjoint} supports with
  increasing means totaling $1000$ points; this data set is used to
  demonstrate the effect of diverse supports amongst the population
  where the PDFs are identical on $\iota$, $G$, $B$, and thus
  $\bar{I}$ versus $\hat{I}$.
\end{dataset}
Each data set will be denoted by $D_i$ where $i$ is
the indexed label of the respective data set.

Finally, to save space, we will demonstrate the TDMI and
non-TDMI-based computations on all the simulated data sets at one
time. We will adhere to the algorithm shown in
Fig. \ref{fig:TDMI_analysis_schematic} when analyzing the real data
sets.

\subsubsection{TDMI-based analysis of the simulated data}

\renewcommand{\arraystretch}{1.5}
\begin{center}
\small
\begin{table*}
\begin{tabular}{|l|l|l|l|l|l|l|l|l|l|l|}
\hline
\multicolumn{11}{|c|}{TDMI-based quantities} \\
\Cline{2pt}{1-11} \hline
\Cline{2pt}{1-11}
Source & $\bar{I}(\tau=1)$ & $\hat{I}(\tau=1)$ & $\bar{B}_{IRP}(\tau=1)$ &
$\hat{B}_{PRP}$ & $\hat{B}_{IRP} (\tau=1)$ & $B_{RP}(\tau=1)$ &
$\mathcal{H}_S (\tau=1)$ &
$\delta \rho (\tau =1)$ & $\delta G (\tau=1)$ & $\delta I(\tau=1)$ \\ \hline \hline
$D_1$ & $0.72$ & --- & $0.008$ & $0.008$ &
$0.008$ & $0$ & $0.99$ & $0$ & $0$ & $0$ \\ \hline \hline
$D_2$ & $0.31$ & --- & $0.012$ & $0.012$ &  $0.012$
& $0$ & $0.96$ & $0$ & $0$ & $0$ \\ \hline \hline
$D_3$ & $0.52$ & $0.37$ & $0.01$ & $0.008$
& $0.007$  & $0.001$ & $0.98$ & $0$& $0.15$ & $0.15$ \\ \hline \hline
$D_4$ & $0.34 \pm 0.07$ & $0.71$ &
$0.18 \pm 0.03$ & $0.013$ &  $0.011$  & $0.002$ & $0.98$ & $0$ &
$\delta I$ & $0.37 \pm 0.07$ \\ \hline \hline
$D_5$ & $0.48 \pm 0.01$ & $0.71$  &
$0.04 \pm 0.01$ & $0.006$ & $0.007$  & $0.001$ & $0.99$ & $0$ &
$\delta I$ & $0.24 \pm 0.01$ \\ \hline \hline
$D_6$ & $0.48 \pm 0.01$
& $1.12$ & $0.04 \pm 0.01$ & $1.12$ & $0.011$  & $1.11$ & $0$ & unknown &
unknown & $0.55 \pm
0.01$ \\ \hline \hline
\end{tabular}
\caption{TDMI results and homogeneity metrics for the simulated data
  sets one through six.}
\label{table:MI_estimates_examples}
\end{table*}
\end{center}

\textbf{Base cases: testing the TDMI-based metrics on
  individuals---}In table \ref{table:MI_estimates_examples} one can
see that both the quadratic and Gauss maps have distinctly different
$I(\tau=1)$ values. Note that the Gauss map has a faster decay in
correlations; for both maps, all correlations in time decay by
$\tau=6$. Further notice that all bias estimation schemes are
essentially identical as expected.  This also implies that
support-variation detecting quantities such as $\mathcal{H}_S$
register no variation in supports.

\textbf{Support dependent, \emph{graph independent} analysis---}To see
how diverse supports are rendered, consider the contrast between
$D_5$ and $D_6$, whose only difference is in the
\emph{location} of the supports. Both of the support-based TDMI based
metrics, $B_{RP}$ and $\mathcal{H}_S$, produced dramatic
representations of the disjoint nature of the supports of data set six
(c.f., table \ref{table:MI_estimates_examples}). Notably, the
difference between both $B_{RP}$ and $\mathcal{H}_S$ on $D_5$
and $D_6$ are near their respective maxima.

\textbf{Graph dependent, support \emph{independent} analysis---}Data
set three, the quadratic-Gauss aggregated data set, has homogeneity in
support in all support-based metrics as can be seen in table
\ref{table:MI_estimates_examples}. In particular, both $\mathcal{H}_S$
and all the random permutation bias estimates are totally unaffected
by the existence of $\bar{\epsilon} \text{ or } \hat{\epsilon} \neq
0$. Furthermore, $\delta I \neq 0$, meaning that the population
averaged TDMI and the TDMI of the aggregated population were
different. In particular, $\bar{I}>\hat{I}$, thus leading to the
conclusion that $\bar{G}>\hat{G}$, which is not surprising given that
when the $\bar{\epsilon}_i = \hat{\epsilon}_i$ for all $i$, it is
reasonable that the $\epsilon$'s register greater though the sum than
the aggregate. In any event, all the TDMI based metrics registered the
diversity in the population of PDFs.

\textbf{Support dependent graph-based analysis---}To begin to see how
support and graph effects intermix, consider $\hat{I}$ for a data set
identical to $D_6$ except where the quadratic data has been
replaced with uniform random numbers, thus yielding data with purely
population location information; denote this data set as $D_6'$. Now,
$\hat{I}(D_6') \approx 1.16 \pm 0.01$, thus comparing $\hat{I}(D_6)$
to $\hat{I}(D_6')$, we notice that the presence of intra-agent
time-based correlation \emph{decreases the population scale TDMI} by a
small but measurable amount --- here $|\hat{I}(D_6) -\hat{I}(D_6')|
\approx 0.04$. Therefore, while nearly all the intra-agent TDMI is
subsumed by the inter-agent TDMI, when there is a presence of both
strong intra-agent information as well as strong inter-agent
information (i.e., highly disjoint supports), $\hat{I}$ will contain
\emph{both} intra-agent and inter-agent components.

What the example in the previous paragraph shows is that deducing the
contribution of the intra-agent and inter-agent components to
$\hat{I}$ will in many cases, be non-trivial. Nevertheless, the use of
metrics that detail the PDF variation can sometimes aid in the
interpretation of $\hat{I}$. First, considering how the heuristic
metrics of PDF variation render the variation in PDFs, note that both
the super sensitive $H_{RA}$ and more robust, less sensitive $V(p)$,
for $D_6$, are about double their values for $D_5$,
even though $D_5$ will yield considerably noisier PDF
estimates. Similarly, the TDMI metrics for PDF variation also render
population diversity; $\delta I$ for $D_6$ is more than twice
$\delta I$ for $D_5$. However, $\delta I$ for $D_6$ has
a slightly more complicated interpretation. In particular, while
$\delta I$ represents the difference between the population and the
individual TDMI, there is likely a non-trivial component of $\bar{I}$
that is a function of sample size.  Thus, $\delta{I}$ is not purely
the difference between the individual and the population TDMI for
unlimited data as it was for $D_3$. Nevertheless, because
$\bar{I} \gg B_E(D_6)$, and $\delta I \gg B_E(D_6)$ we know that
$\hat{I}$ has components of both individual and population scale
TDMI. In fact, considering $|\hat{I}(D_5) - \hat{I}(D_6)| \approx
0.41$ versus $|\hat{I}(D_5) - \hat{I}(D_6')| \approx 0.44$, one can
see that for this case, the TDMI whose source is in the population
dominates; presumably if the supports for $D_6$ were nearly
overlapping instead of disjoint, $|\hat{I}(D_5) - \hat{I}(D_6)|$ would
be much closer to zero. While it is unusual to be able to compare
identical, stationary systems with differing supports, this analysis
does suggest that calculating $\hat{I}$ for the raw data and for the
data with normalized supports may be useful for determining the
proportion of $\hat{I}$ that is due the diversity of the supports.

\subsubsection{Non-TDMI-based analysis of the simulated data}

\renewcommand{\arraystretch}{1.5}
\begin{center}
\small
\begin{table*}
\begin{tabular}{|l||l|l|l|l|l|l|l|}
\hline
\multicolumn{8}{|c|}{non-TDMI-based population diversity metrics} \\
\Cline{2pt}{1-8} \hline
\Cline{2pt}{1-8}
Source & $H(\bar{x})$ & Var($n_i$) & $s_{min} \pm V_{s_{min}}$ &
$s_{max} \pm V_{s_{max}}$ &  $|S| \pm V_{|S|}$ & $H_{RA}$ &  $V_{\bar{\mathcal{S}}}(p)$ \\
\hline \hline
D1 & $0$ & $0$ & $0.0001$ & $0.999$ & $0.9989$ & $0$ &
$0$ \\ \hline \hline
D2 & $0$ & $0$ & $0.0002$ & $0.9998$  & $0.9997$ & $0$ & $0$ \\ \hline \hline
D3 & $0$ & $0$ & $0.0002 \pm 0.0003$ &
$0.9989 \pm 0.0015$ & $0.9987 \pm 0.0018$ & $0.16$ & $0.09$ \\ \hline
D4 & $0$ & $0$ & $0.02 \pm 0.02$ & $0.996 \pm 0.006$ & $0.98 \pm 0.03$
& $0.9$ & $0.39$\\ \hline \hline
D5 & $0$ & $0$ & $0.001 \pm 0.002$ &
$0.9997 \pm 0.0006$ & $0.998 \pm 0.003$ & $0.37$ & $0.13$ \\ \hline \hline
D6 & $0$ & $0$ & $5.5
\pm 3$ & $6.5 \pm 3$ & $0.997 \pm 0.004$ & $0.68$ & $0.32$ \\ \hline \hline
\end{tabular}
\caption{Heuristic homogeneity metrics for the simulated data sets one
though six.}
\label{table:homo_metrics_examples}
\end{table*}
\end{center}

\textbf{Base cases: testing the non-TDMI metrics on
  individuals---}Begin by considering $D_1$ and $D_2$, both of
which represent only a single individual. Both cases are well defined
in $p$ (c.f., Fig. \ref{fig:f_diff_agg}), and have supports whose
lengths, $|S|$, and boundaries, $s_{min}$, $s_{max}$, are well
resolved and within their expected ranges (c.f., table
\ref{table:homo_metrics_examples}).

\textbf{Support dependent, \emph{graph independent} analysis---}To see
how variations in the supports are rendered, consider the contrast
between $D_5$ and $D_6$, whose only difference is in the
\emph{location} of the supports. Focusing on $D_6$, variation
in the support shows up in the heuristic metrics $s_{min}$, $s_{max}$,
$|S|$, and especially in the variance of $s_{min}$ and $s_{max}$.

\textbf{Graph dependent, support \emph{independent} analysis---}Data
set three, the quadratic-Gauss aggregated data set, has homogeneity in
support in all support-based metrics as can be seen in tables
\ref{table:homo_metrics_examples} as expected. In contrast, both of
the heuristic metrics designed to detect variation in PDFs registered
as non-zero, meaning they detected variation in the PDFs.  Moreover,
the $l_1$-like diagnostic, $H_{RA}$ was more sensitive than the
variance based metric,  $V_{\bar{\mathcal{S}}}(p)$, as expected.

\textbf{Support dependent graph-based analysis---}None of the examples
mix graph and support effects simultaneously by design.

\subsubsection{Quantifying small sample-size effects}

To form a baseline of small sample size effects for both real data
applications and the support-based effects, we focus on comparing and
constraining results for $D_4$ and $D_5$, the quadratic map
data sets with $50$ sets of $20$ points, and $10$ sets of $100$
points.

\textbf{Small sample size effects on non-TDMI-based support analysis
  metrics---}The heuristic metrics of support diversity show
homogeneity in support. However, it is important to note that the
invariant density of the quadratic map has most of its mass at the end
points, and thus may represent the best case scenario for support
based metrics on small data sets.

\textbf{Small sample size effects on TDMI-based support analysis
  metrics---}The TDMI based metrics of support diversity show
homogeneity of support, although the individual-wise random
perturbation for the random case ($B_{IRP}$) is rather high,
especially for the $20$ point data sets, as one might expect. However,
we hypothesize that the primary reason why $B_{IRP}$ is so high for
the $20$ point data sets is that, upon randomly permuting any data
set, the average $\tau$ will be the length of the data set over $3$,
in this case, $\frac{20}{3} < 7$. Thus, for very short data sets, it
can be difficult to approximate the estimator bias using only the
random permutation method \cite{estimation_of_MI_EHR}.

\textbf{Small sample size effects on non-TDMI-based graph analysis
  metrics---}In contrast to the support-based effects, the
heuristic-based PDF variability metrics register \emph{substantial}
diversity among the PDFs $D_4$ and $D_5$, effects that are
entirely a function of small sample sizes. These results are not
surprising given that there will be great variance in the PDF estimate
of a quadratic time-series with only $20$ points.

\textbf{Small sample size effects on TDMI-based graph analysis
  metrics---}The small sample size situation highlights both the
difference between $\bar{I}$ and $\hat{I}$ and also displays the
motivation for why one would want to estimate $\hat{I}$. The
\emph{average} based TDMI results for both $D_4$ and $D_5$
\emph{do not} approximate the $1000$ point analogs; and moreover, the
addition of more \emph{sets} of data with similar lengths will not
help the $\bar{I}$ to converge to the higher point analog but rather
decrease the variance in the mean $\bar{I}$ value. Thus, the desired
meaning of $\bar{I}$ is, in a sense, a precision/accuracy type
problem; adding more $20$ point data sets will make the estimate of
$\bar{I}$ more precise, but not necessarily more accurate. That said,
accuracy is always defined relative to a target; there is likely less
TDMI in the $20$ point data set because there is considerably less
time-based information in a $20$ or $100$ point data set than in a
$1000$ point data set.  Therefore, while adding more data sets will
not aid in convergence to the infinite point analog, the infinite
point analog may not be right target to be aiming for with $20$ point
data sets. In contrast, the \emph{aggregated} data sets produce a TDMI
equivalent to the $1000$ point analog, thus inducing a $\delta
I$. Moreover, adding points to the aggregated data set will help with
convergence to $I(\tau=1)$ for infinitely long data strings.

\textbf{Interpreting $\delta I$ when individual elements have few
  pairs of points---}The existence of $\delta I$ for $D_4$ and $D_5$
introduces a form of divergence from $I(\tau=1, N=\infty)$ that is not
quite a bias (either estimator or non-estimator); the ``true'' amount
of information in a data string of length $20$ is fundamentally
different from the ``true'' amount of information in a data string of
length $N=\infty$ --- thus $\delta I$ can also exist due to finite
sample size effects. Or, said more quantitatively, $\bar{I}$, even for
an unlimited collection of $100$ point data strings, will never be
within estimator bias or any other kind of bias, of $I(\tau=1,
N=\infty)$ because $I(\tau=1, N=\infty) \sim 0.72$ while $I(\tau=1,
N=20) \approx 0.48 \pm 0.1$. What this means for $\hat{I}$ is that,
unless the aggregated data sets are homogeneous enough in their
time-dependent correlation structure, $\hat{I}$ will likely represent
\emph{population distribution} information, as $\bar{I}$ would
represent the upper bound on time-correlation based information
present in each data string. Often the composition of most real world data
streams can be difficult to infer; and moreover, it can be a
non-trivial problem to discern whether $\bar{I}$ or $\hat{I}$ most
faithfully represent a population or individual effects. For instance, in
Ref. \cite{stat_dyn_diurnal_correlation_short}, the authors claim both
the presence of time-correlation information and population-based
time-correlation being simultaneously present. Usually a careful
analysis of the population composition of the $\delta t$ bins will
help rectify this difficulty.

\subsection{Real data examples: glucose values for $100$ densely
  sampled individuals versus $20,000$ random individuals}

\begin{figure}
    \subfigure[Individual PDF estimates for the $100$ patients with
      the largest record]
    {
      \epsfig{file=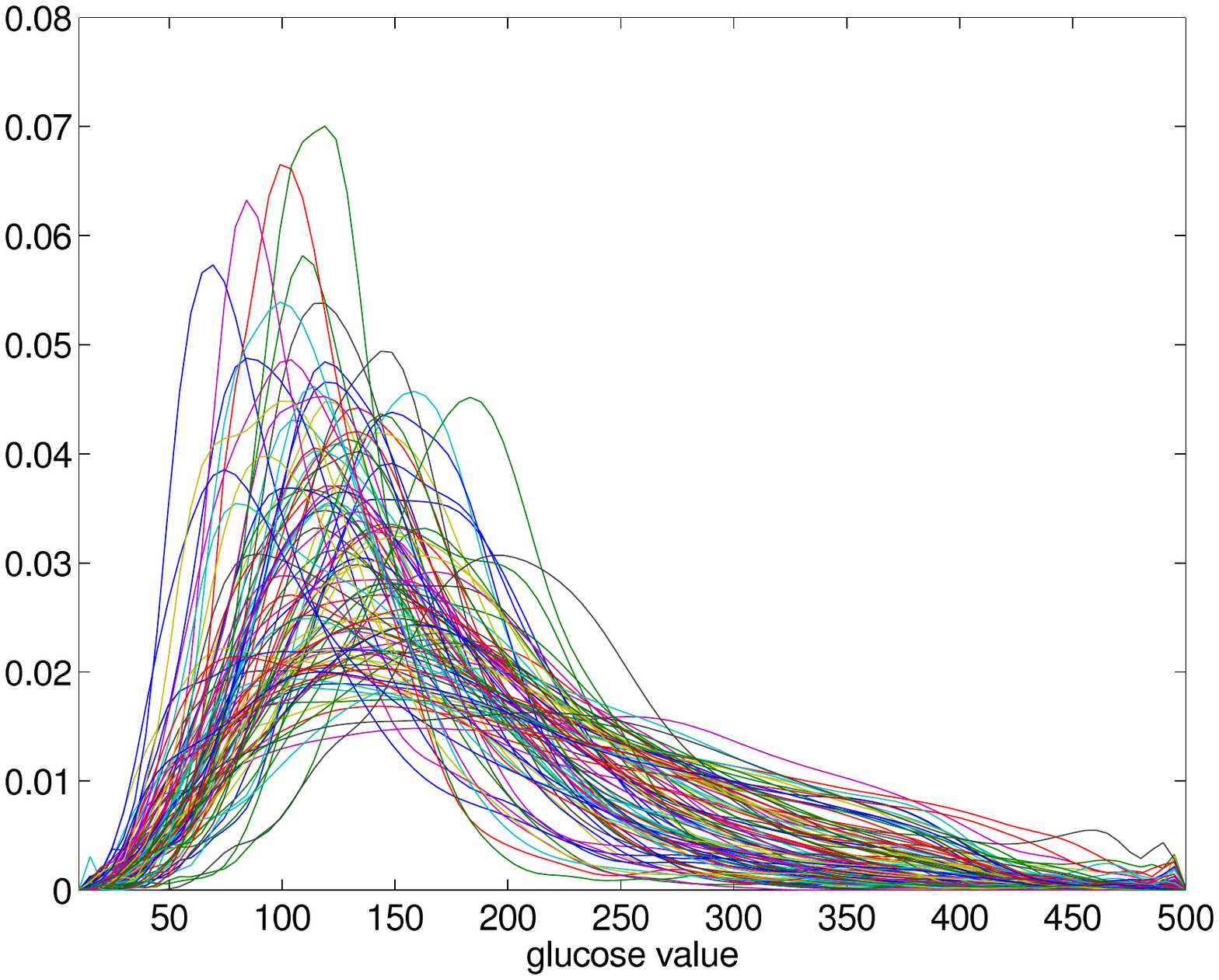, height=6cm}
      \label{figure:patient_pdf_compare_a}
    }
    \subfigure[Individual PDF estimates for the $20,000$ random patients]
    {
      \epsfig{file=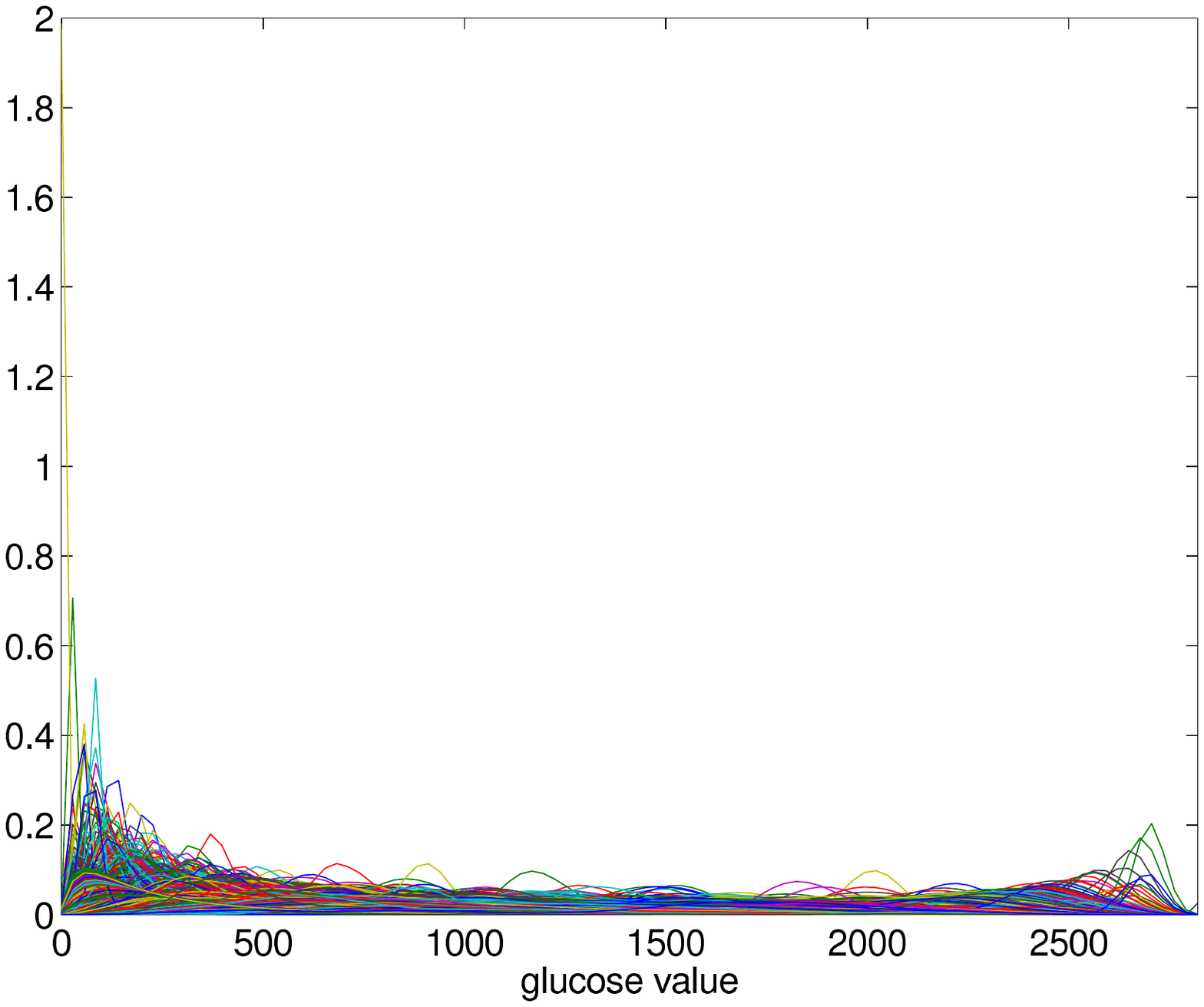, height=6cm}
      \label{figure:patient_pdf_compare_b}
    }
    \subfigure[Aggregated population PDF comparison]
    {
      \epsfig{file=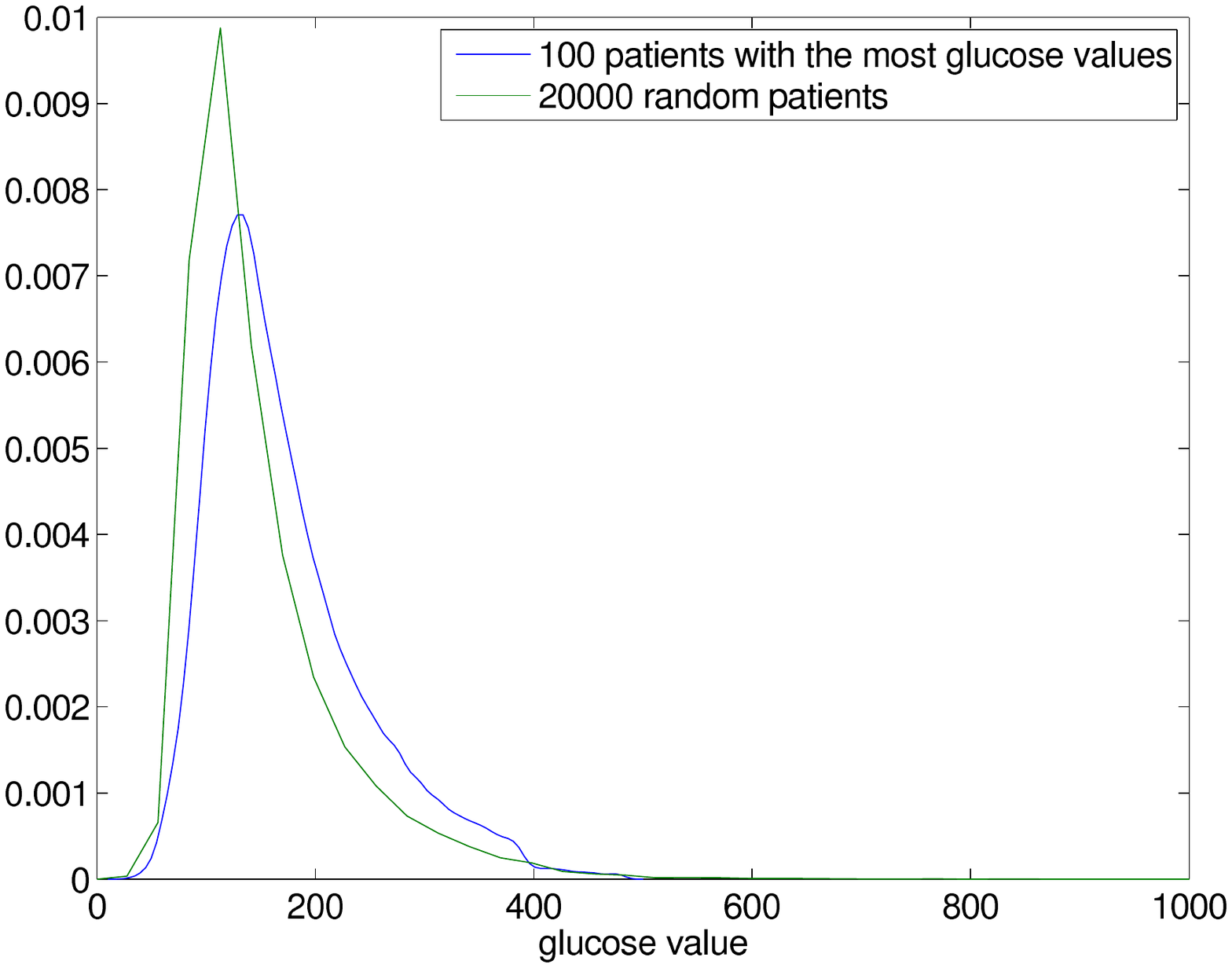, height=6cm}
      \label{figure:patient_pdf_compare_c}
    }
    \caption{PDFs of glucose measurements for individuals within a
      population and for a population for two data sets, the $100$
      patients with the largest records and $20,000$ random patients}
    \label{figure:patient_pdf_compare}
\end{figure}

We now move on to applying the insights and techniques of the previous
sections to real data.  In particular, we will consider two data sets
that contain different populations of patients from the CUMC data
repository. More specifically, the data sets include:
\begin{dataset}
  a collection of the $100$ patients with the most glucose
  measurements in the database, ranging from $\sim 4000$ to $\sim
  1500$ measurements per patient;
\end{dataset}
\begin{dataset}
  a collection of $20,000$ random patients with at least $2$ glucose
  measurements from among the $800,000$ patients with glucose values.
\end{dataset}
To visualize these populations, consider
Fig. \ref{figure:patient_pdf_compare} where the normalized PDFs for
each individual for each population and the PDF of the overall
populations are plotted. While the population-wide PDFs, shown in
Fig. \ref{figure:patient_pdf_compare_c} are not wildly different, the
relative diversity within the two populations, as shown in
Figs. \ref{figure:patient_pdf_compare_a} and
\ref{figure:patient_pdf_compare_b}, is dramatic. The motivation for
choosing $D_7$ is that, for this set, because each patient has
\emph{at least} $1000$ lab values, both $\bar{I}$ and $\hat{I}$ are
calculable. Moreover, the authors hypothesize that patients with so
many glucose values are more likely to represent a more homogeneous
population compared with the population at large. Given the makeup of
$D_7$, $D_8$ represents not only a contrast to $D_7$ in that $D_8$ is
a snapshot of the entire population, but $D_8$ also represents a
pathologically difficult situation data-wise --- very few patients
have more than $100$ glucose values, and the set of possible causes
for the existence of a glucose measurement is extremely large (or
broad). Thus, not only will $\bar{I}$ be difficult to calculate for
$D_8$ (most patients won't have enough data to generate a PDF
estimate), but there is likely tremendous and differing diversity
amongst the patients actually included in the estimates of $\bar{I}$
and $\hat{I}$.

Finally, note that in contrast to the previous analysis of simulated
data, we will present the TDMI results first, followed by an analysis
using the non-TDMI metrics to verify the TDMI results. The point of
this ordering is to demonstrate the TDMI infrastructure without
hindsight knowledge.


\begin{figure}
    \subfigure[Comparisons of support minima, maxima, and length for
      the two populations]
    {
      \epsfig{file=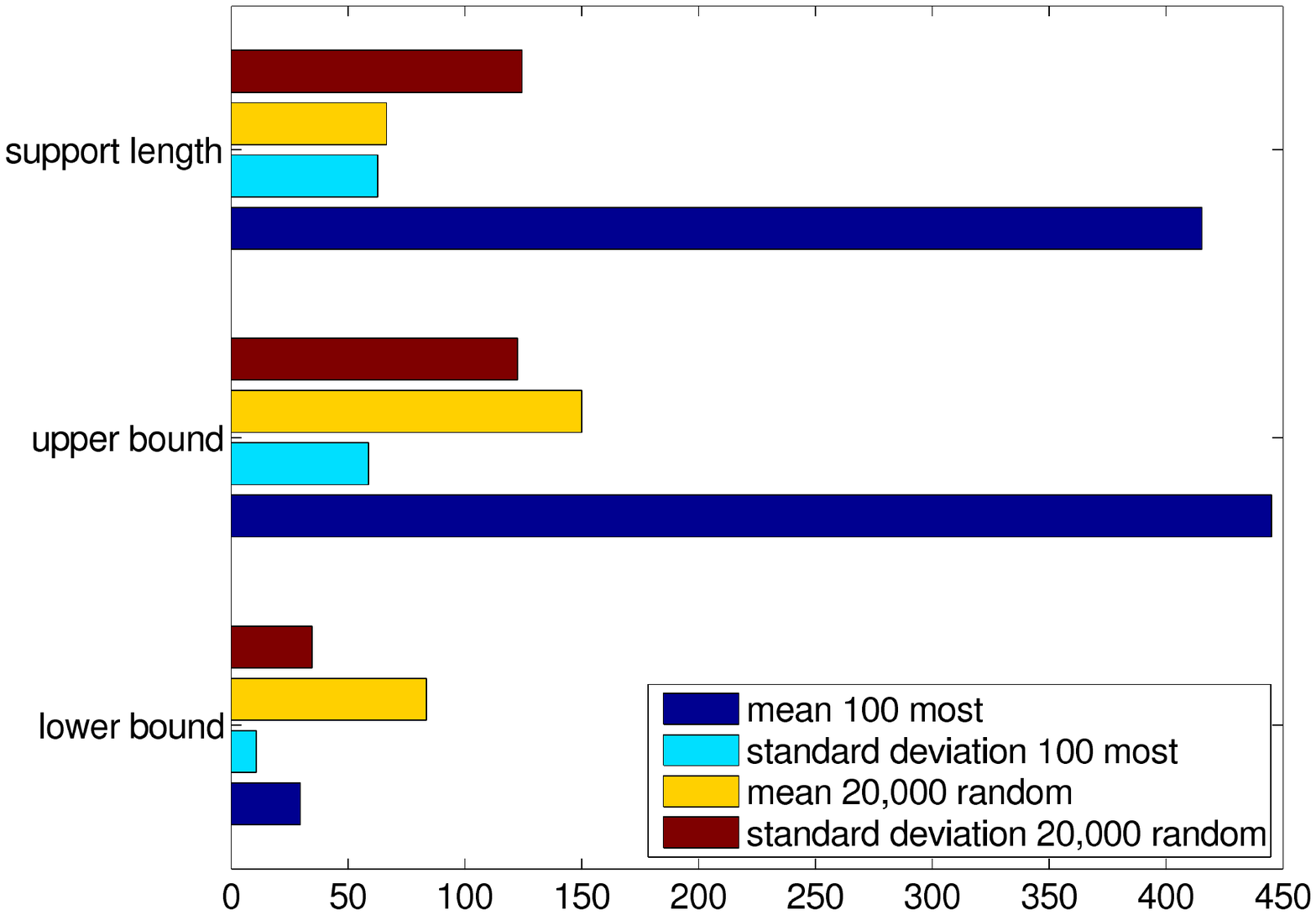, height=6cm}
      \label{figure:patient_compare_a}
    }
    \subfigure[Comparisons of population maxima minus the population
      minima for the two populations]
    {
      \epsfig{file=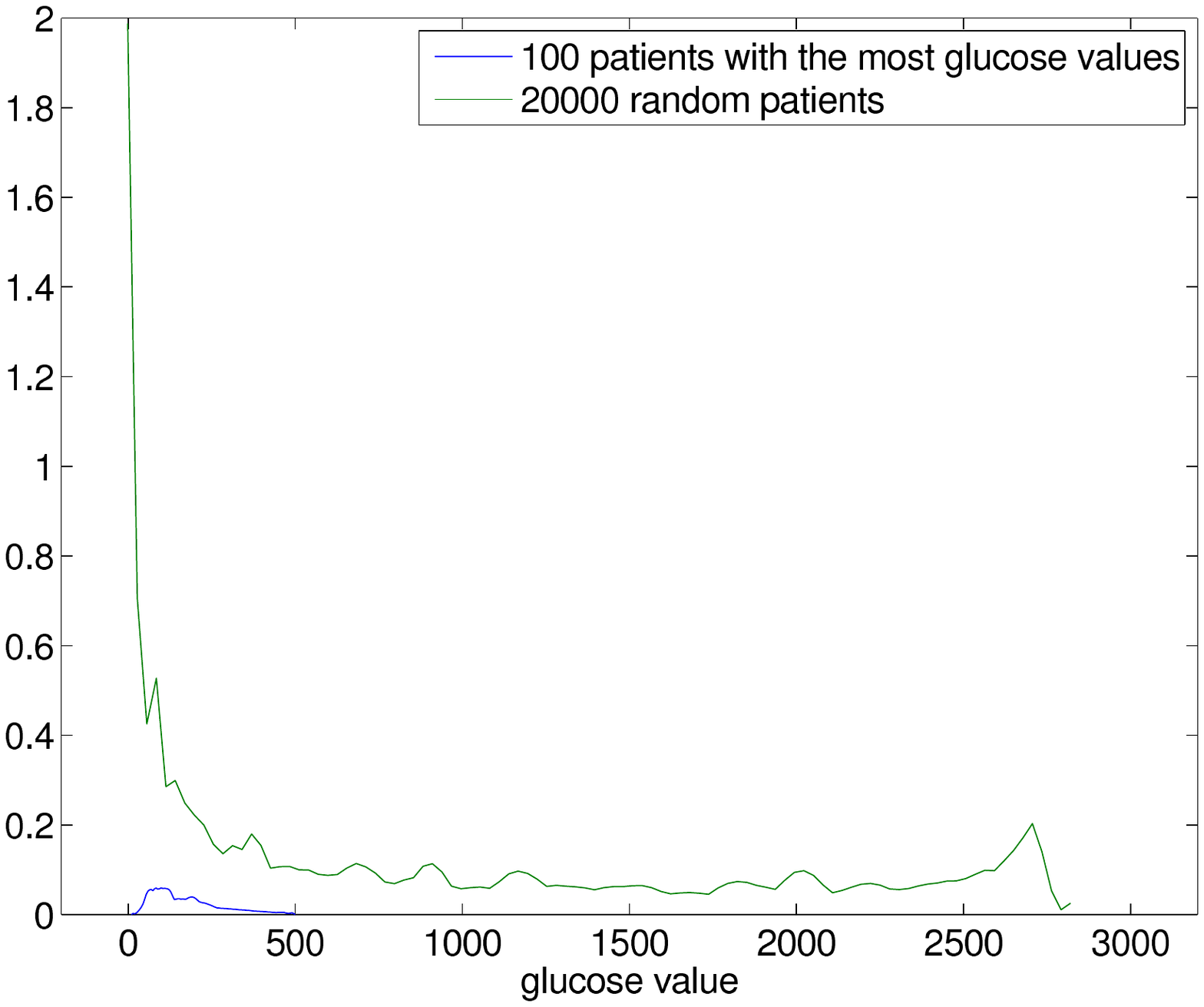, height=6cm}
      \label{figure:patient_compare_b}
    }
    \subfigure[Comparisons of the standard deviation of the PDF graphs for the two populations]
    {
      \epsfig{file=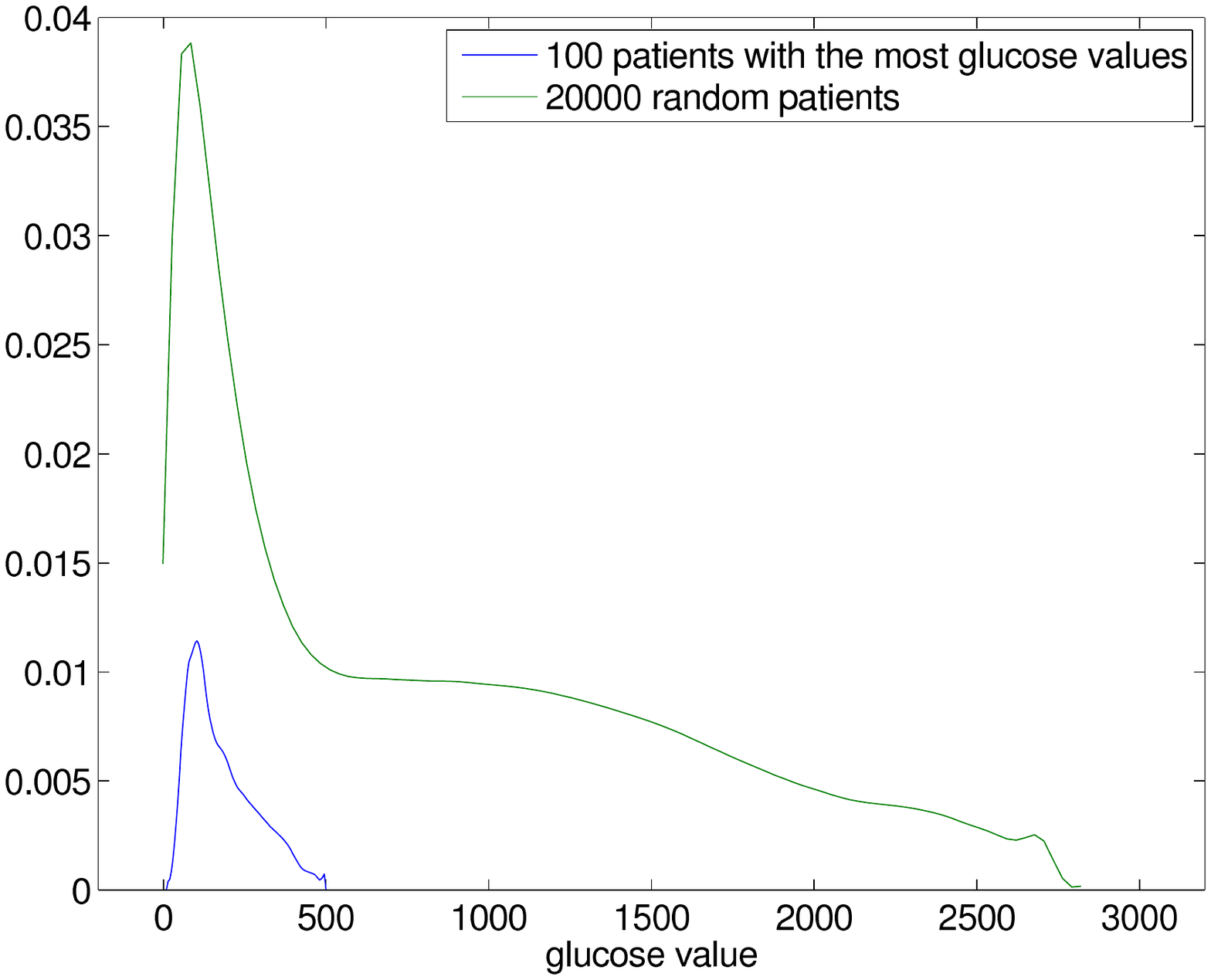, height=6cm}
      \label{figure:patient_compare_c}
    }
    \caption{Comparisons of the supports, and PDF graph variations for
      two data sets, the $100$ patients with the largest records and
      $5000$ random patients}
    \label{figure:patient_compare}
\end{figure}

\subsubsection{TDMI-based analysis for data set 7, the well measured population}

\renewcommand{\arraystretch}{1.5}
\begin{center}
\small
\begin{table*}
\begin{tabular}{|l|l|l|l|l|l|l|l|l|l|l|}
\hline
\multicolumn{11}{|c|}{TDMI-based quantities for the $\delta t = 6$ hrs
time separation} \\
\Cline{2pt}{1-11} \hline
\Cline{2pt}{1-11} 
Source & $\bar{I}$ &  $\hat{I}$ & $\delta I $ &
$\bar{B}_{PRP}$ & $\hat{B}_{IRP}$  & $\hat{B}_{PRP}$ &
$B_{RP}$ & $\mathcal{H}_S$ & $H_{\Theta}$ & $N_{min}$ \\ \hline \hline
$D_7$ & $0.64 \pm 0.03$ & $0.22$ & $0.42 \pm 0.03$ & $0.02 \pm 0.01$ &
$0.02 \pm 0.005$ & $0.001 \pm 0.0005$ & $\sim \hat{B}_{IRP}$ & $1 \pm
0.0005$ & $0.31$ & $470$ \\ \hline \hline
$D_8$ & $0.29 \pm 0.16$ & $0.38$ & $0.09 \pm 0.37$ & $0.2 \pm 0.2$ &
$0.08 \pm 0.005$ & $0.006 \pm 0.0005$ & $\sim \hat{B}_{IRP}$ & $1 \pm 0.02$
& $0.003$ & $1$ \\ \hline \hline
\end{tabular}
\caption{TDMI results and homogeneity metrics for the real patient data
  sets seven and eight; note all $\delta t$ times are in hours.}
\label{table:MI_estimates_patients_6}
\end{table*}
\end{center}

\renewcommand{\arraystretch}{1.5}
\begin{center}
\small
\begin{table*}
\begin{tabular}{|l|l|l|l|l|l|l|l|l|l|l|}
\hline
\multicolumn{11}{|c|}{TDMI-based quantities for the $\delta t = 24$ hrs
time separation} \\
\Cline{2pt}{1-11} \hline
\Cline{2pt}{1-11}
Source & $\bar{I}$ &  $\hat{I}$ & $\delta I $ &
$\bar{B}_{PRP}$ & $\hat{B}_{IRP}$  & $\hat{B}_{PRP}$ &
$B_{RP}$ & $\mathcal{H}_S$ & $H_{\Theta}$ & $N_{min}$ \\ \hline \hline
$D_7$ & $0.093 \pm 0.06$ & $0.077$ & $0.016 \pm 0.06$ & $0.02 \pm 0.01$ & $0.02 \pm
0.005$ & $0.001 \pm 0.0005$ & $\sim \hat{B}_{IRP}$ & $0.99 \pm 0.01$ & $0.33$ & $479$ \\ \hline \hline
$D_8$ & $0.21 \pm 0.15$ & $0.17$ & $0.04 \pm 0.15$ & $0.3 \pm 0.2$ & $0.07 \pm 0.01$
& $0.005 \pm 0.001$ & $\sim \hat{B}_{IRP}$ & $0.97 \pm 0.001$ &
$0.005$ & $1$ \\ \hline \hline
\end{tabular}
\caption{TDMI results and homogeneity metrics for the real patient data
  sets seven and eight; note all $\delta t$ times are in hours.}
\label{table:MI_estimates_patients_24}
\end{table*}
\end{center}


\textbf{Analysis of the $\delta t = 6$ hrs time separation using the
  algorithm in Fig. \ref{fig:TDMI_analysis_schematic}---}First,
considering table \ref{table:MI_estimates_patients_6}, note that for
$D_7$ with a $\delta t=6$hrs, we are able to estimate $\bar{I}$, and
thus $\delta I$ because $N_{min}(6 hrs) > 100$. Next, note that
$\delta I(6 hrs)$ is considerably above $B_{IRP}(6 hrs)$, meaning that the
population is on the time-scale of $6$ hours is
heterogeneous. Moreover, both $\bar{I}(6 hrs)$ and $\hat{I}(6 hrs)$ are greater than
zero, meaning that there is TDMI present in individuals and the
aggregated population. To determine the nature of heterogeneity,
further consider the support-based metric; $\mathcal{H}_S(6 hrs) \sim 1$
points to the population having uniformity in supports or ranges
($B_{RP}(6 hrs) \approx B_{IRP}(6 hrs)$ which corroborates this
conclusion). Finally, the entire population is reasonably represented
for $\delta t =6hrs$ as confirmed by the fact that $N_{min}(6 hrs) \sim 500$
and $H_{\Theta}(6 hrs) \gg 0$. Thus, the concluding interpretation is as
follows: the population is heterogeneous on the $\delta t = 6$hrs time
scale; the heterogeneity in the population is in the graphs not the
supports (or the normalizations; there is diverse but present temporal
correlation among the population (i.e., the TDMI is not due to the
population aggregation, but exists because of the individuals); and
the entire population is well represented in the TDMI-based
quantities.

\textbf{Analysis of the $\delta t = 24$ hrs time separation using the
  algorithm in Fig. \ref{fig:TDMI_analysis_schematic}---}First,
considering table \ref{table:MI_estimates_patients_24}, note that for
$D_7$ with a $\delta t=24$hrs, we are able to estimate $\bar{I}$, and
thus $\delta I$ because $N_{min}(24 hrs) > 100$. Next, note that
$\delta I(24 hrs)$ is within the error bars of zero (e.g., below
$B_{IRP}(24 hrs) $), meaning that the population is on the time-scale of $24$
hours is \emph{homogeneous}. Moreover, both $\bar{I}(24 hrs) $ and $\hat{I}(24 hrs) $
are greater than zero, meaning that there is TDMI present in
individuals and the aggregated population. To determine the nature of
heterogeneity, further consider the support-based metric;
$\mathcal{H}_S(24 hrs)  \sim 1$ points to the population having uniformity in
supports or ranges ($B_{RP}(24 hrs)  \approx B_{IRP}(24 hrs) $ which corroborates this
conclusion). Finally, the entire population is reasonably represented
for $\delta t =24hrs$ as confirmed by the fact that $N_{min}(24 hrs)  ~ 500$
and $H_{\Theta}(24 hrs)  \gg 0$. Thus, the concluding interpretation is as
follows: the population is homogeneous on the $\delta t = 24$hrs time
scale; there is present temporal correlation among the population
(i.e., the TDMI is not due to the population aggregation, but exists
because of the individuals); and the entire population is well
represented in the TDMI-based quantities.

\renewcommand{\arraystretch}{1.5}
\begin{center}
\small
\begin{table}
\begin{tabular}{|l|l|l|}
\hline
\multicolumn{3}{|c|}{time independent TDMI-based quantities} \\
\Cline{2pt}{1-3} \hline
\Cline{2pt}{1-3}
Source & $\bar{h}$ & $\hat{h}$  \\ \hline \hline
$D_7$ &  $1.39 \pm 0.07$ & $2.12$   \\ \hline \hline
$D_8$ &  $0.8 \pm 0.22$ &  $2.05$ \\ \hline \hline
\end{tabular}
\caption{Time \emph{independent} TDMI results  for the real patient data
  sets seven and eight..}
\label{table:MI_estimates_patients_time_independent}
\end{table}
\end{center}

\textbf{Analysis independent of time---}Considering the entropy
calculations in table
\ref{table:MI_estimates_patients_time_independent}, $D_7$ renders some
heterogeneity because the difference between $\bar{h}$ and $\hat{h}$
is non-zero. Nevertheless, as we will see for $D_8$, an entropy
difference of $0.73$, which is about half the magnitude of $\bar{h}$,
would argue that the \emph{static} information theoretic
interpretation of the population is of relative homogeneity.

\textbf{Sample size issues---}There were no sample size issues with
respect to either $\delta t$ time separations studied; in both cases,
$N_{min}$ was well over $100$, and thus all PDFs and their respective
biases could be accurately estimated. In fact, careful analysis of the
population make-up in each $\delta t$ between $6$hrs and $56$hrs
revealed that the proportionally of each individual remained
relatively constant. Finally Fig.  \ref{fig:ehr_ave_v_agg}, where both
the TDMI estimated using both KDE and histogram estimation schemes are
shown, confirms the lack of any small sample size effects because both
estimation schemes are essentially equal.

\subsubsection{non-TDMI-based analysis for data set 7, the well measured population}

\renewcommand{\arraystretch}{1.5}
\begin{center}
\small
\begin{table*}
\begin{tabular}{|l|l|l|l|l|l|l|l|}
\hline
\multicolumn{8}{|c|}{non-TDMI-based analysis metrics} \\
\Cline{2pt}{1-8} \hline
\Cline{2pt}{1-8}
Source & $H(\bar{x})$ & Var($n_i$) & $s_{min} \pm V_{s_{min}}$ &
$s_{max} \pm V_{s_{max}}$ &  $|S| \pm V_{|S|}$ & $H_{RA}$ & $V_{\bar{\mathcal{S}}}(p)$ \\ \hline
$D_7$ & $1.042$ & $463.7$ &  $29.7 \pm  10.7$ & $445.0
\pm58.8$ & $415.4 \pm 62.7$ & $0.898$ & $0.432$ \\ \hline
$D_8$ & $30$ & $55$ & $84 \pm 35$ & $150
\pm 122$ & $66 \pm 125$ & $1$ & $0.90$ \\ \hline \hline
\end{tabular}
\caption{Heuristic homogeneity metrics for the real patient data sets
  seven and eight.}
\label{table:homo_metrics_patients}
\end{table*}
\end{center}

\textbf{Non-TDMI support-based analysis---}To verify the TDMI-based
results, begin by observing that heuristic metric that quantifies
variation in the supports, $H(\bar{X}) \approx 1$, which is considered
small. Thus, while there is some diversity among how the patients were
measured, variation how patients are measured is small. This claim is
also justified by the fact that the variance in the number of points
contributed, per patient, to the $\delta t = 6 hrs$ bin, $Var(n_i)$,
is small. Finally, the variance in $s_{min}$, $s_{max}$ and $|S|$ is
small compared to the respective values (c.f.,
Fig. \ref{figure:patient_compare_a}). Because these are
\emph{time-independent} measures of the support, and because adding
the temporal aspect of the analysis only makes the data set smaller,
it is likely that the TDMI analysis of the homogeneity of support are
correct.

\textbf{Non-TDMI graph-based analysis---}The most sensitive PDF
variation metric, $H_{RA}$ points to a relatively diverse population,
while the less sensitive PDF variation metric
$V_{\bar{\mathcal{S}}}(p)$, based on the standard deviation of the
\emph{distribution} of PDFs, points to a relatively homogeneous, yet
not totally homogeneous population. Figure
\ref{figure:patient_compare} confirms this analysis visually. The
maxima minus the minima, which, when integrated is essentially
$H_{RA}$, shown in Fig. \ref{figure:patient_compare_b}, can be seen to
be relatively large, thus making $H_{RA}$ render diversity. In
contrast, the variance in the graphs of the PDFs, shown in
Fig. \ref{figure:patient_compare_c}, is seen as relatively small for
$D_7$, thus making $V_{\bar{\mathcal{S}}}(p)$ render relative
homogeneity. It is important to note, however, that
$V_{\bar{\mathcal{S}}}(p)$, which is independent of time, does not
detail the fact that the population has diverse predictive information
for time periods less than $6$ hours; this is an important distinction
to make as it implies that prediction can vary with time despite the
overall distribution of physiological variables. Finally, both the
TDMI and the heuristic analysis conclude that the population is
homogeneous in supports and in the long term (i.e., independent of
time), the population is homogeneous; this is because $\delta I \sim
0$ for $\delta t > 12$ hrs and $V_{\bar{\mathcal{S}}}(p)$ is small.


\subsubsection{TDMI-based analysis for data set 8, the random (less
  well measured) population}

\textbf{Analysis of the $\delta t = 6$ hrs time separation using the
  algorithm in Fig. \ref{fig:TDMI_analysis_schematic}---}First,
considering table \ref{table:MI_estimates_patients_6}, note that for
$D_8$ with a $\delta t=6$hrs, we are \emph{not} really able estimate
$\bar{I}(6 hrs)$ because $N_{min}(6 hrs) = 1$. To interpret $\hat{I}(6 hrs)$, we
consider the support-based metric; $\mathcal{H}_S (6 hrs) \sim 1$ which points
to the population, \emph{which was filtered and has time points
  separated by 6 hours}, having uniformity in supports or ranges
($B_{RP}(6 hrs) \approx B_{IRP}(6 hrs)$ which corroborates this conclusion). To give
intuition to the graph-based variation, consider
$V_{\bar{\mathcal{S}}}(p)$ (table \ref{table:homo_metrics_patients}),
which implies a somewhat diverse population. Moreover,
$V_{\bar{\mathcal{S}}}(p)$ for $D_8$ is \emph{twice that} of $D_7$,
implying that the population in $D_8$ is more diverse than that of
$D_7$. Moving beyond the algorithm shown in
Fig. \ref{fig:TDMI_analysis_schematic}, we did estimate $\bar{I}(6
hrs)$ and thus, $\delta I(6 hrs)$, only including individuals with
enough points to estimate $I$. Based on this restricted version of
$\delta I(6 hrs)$, the population appears to be homogeneous.
Nevertheless, both the restricted $\bar{I}(6 hrs)$ and $\hat{I}(6
hrs)$ are greater than zero, meaning that there is TDMI present in
individuals and the aggregated population. This means that there is an
apparent contradiction; the \emph{restricted} $\delta I(6 hrs)$ implies a
population that is somewhat homogeneous/heterogeneous while
$V_{\bar{\mathcal{S}}}(p)$ implies a heterogeneous population. This
contradiction is resolved by recalling that $V_{\bar{\mathcal{S}}}(p)$
is calculated on the \emph{entire, non-filtered population} and is
\emph{independent of time} and will overestimate graphic diversity,
while $\delta I$ is overly restricted and will underestimate
diversity. This interpretation will be substantiated further in
sections \ref{sec:dynamic_TDMI} and \ref{sec:billing}. Finally, the
overall population is \emph{poorly} represented for $\delta t =6hrs$
as confirmed by the fact that $N_{min}(6 hrs) = 1$ and $H_{\Theta}(6
hrs) \approx 0$. In fact, for $D_8$, we know that $63 \%$ of the
patients ($12,763$) have \emph{no points} in the $\delta t = 6 hrs$
bin, and only $12 \%$ ($2,400$) of the patients have ten or more
points in the $\delta t = 6 hrs$ bin. Thus, the concluding
interpretation is as follows: the population is homogeneous on the
$\delta t = 6$hrs time scale up to what is resolvable by $\delta I (6
hrs)$; the represented population has relatively uniform supports;
there is diverse but present temporal correlation among the population
(i.e., the TDMI is not due to the population aggregation, but exists
because of the individuals); the population has diversity relative to
their time-independent graphs, but this graph diversity may not
reflect the graph diversity of the represented population (i.e., the
population used to estimate the TDMI-based quantities); the overall
population of patients is poorly represented in the TDMI-based
diagnostics; and finally the overall population of $20,000$ patients
is diverse, but the patients that have enough data to estimate the
TDMI on time-scales of $\delta t \leq 48 hrs$ (i.e., the represented
population), which represents a strongly filtered subpopulation, is
relatively homogeneous in predictive information \emph{regardless of
  $\delta t$}.

\textbf{Analysis of the $\delta t = 24$ hrs time separation using the
  algorithm in Fig. \ref{fig:TDMI_analysis_schematic}---}Considering
table \ref{table:MI_estimates_patients_24} (and later,
Fig. \ref{fig:ehr_ave_v_agg_b}), the analysis of the TDMI diagnostics
for $\delta t = 24$hrs is essentially identical to $\delta t=6$hrs
case. Even representative population for both the $\delta t =6$ and
$24$hrs bins is essentially identical down to the individual
proportional contributions to the aggregated data set. Thus, the key
observation here is the difference between $D_7$ and $D_8$; $D_7$
registered heterogeneity at $\delta t=6$hrs and homogeneity at $\delta
t=24$hrs whereas $D_8$ does not render a $\delta t$ dependence in the
TDMI-based diagnostics.

\textbf{Analysis independent of time---}Considering the entropy
calculations in table
\ref{table:MI_estimates_patients_time_independent}, $D_8$ renders
heterogeneity because the difference between $\bar{h}$ and $\hat{h}$
is non-zero. In particular, compared to the entropy differences for
$D_7$, the $D_8$ has an entropy difference of $\sim 1.25$, which is
substantially \emph{larger} in magnitude than $\bar{h}$. Thus the
\emph{static} information theoretic interpretation of the population
in $D_8$, which includes all patients (there is not filtering effect),
is of heterogeneity.

\textbf{Sample size issues---}There are three sample size issues
present in the TDMI analysis of $D_8$, the poor representation of the
overall population, the inability to estimate $I$ for every
representative member of the population, and the overall small sample
size and bandwidth/normalization issues. The first issue implies that
the probability mass used to estimate the PDFs comes from a \emph{very
  small subset of the population}; e.g., only $12 \%$ of the
population has $10$ or more points in the $\delta t = 6 hrs$
bin. Thus, the restricted (i.e., filtered) population is likely
substantially more homogeneous than the overall population, and the
TDMI analysis cannot be said to represent the overall
population. Relative to the second issue, since $N_{min} = 1$ (for
both $\delta t = 6$ and $24$ hrs), $\bar{I}(\delta t)$ is
representative of a smaller population than $\hat{I}(\delta
t)$. Finally the third issue, small sample size effects, can be seen
in the large difference (about a factor of $2$) between the KDE and
histogram estimator based TDMI values seen in
Fig. \ref{fig:ehr_ave_v_agg}.

\subsubsection{non-TDMI-based analysis for data set 8, the random
  (less well measured) population}

\textbf{Non-TDMI support-based analysis---}Begin by noticing that
there is considerable diversity in how the $20,000$ patients are
measured, as can be seen in $H(\bar{X}) \approx 30$, which is $30$
times larger $H(\bar{X})$ for $D_7$.  Considering this in conjunction
with $Var(n_i) \approx 50$ for $D_8$, which is much smaller than
$Var(n_i)$ for $D_7$, implies that very few of the patients have many
points.  Said differently, the reason why $Var(n_i)$ is relatively
small compared to $H(\bar{X})$ is that $n_i$ is bounded from below by
$0$ and is never very large for any member of $D_8$. That this is the
fact is reflected in variance in $s_{min}$, $s_{max}$ and $|S|$, which
is large (on the order of, or greater than) the values of $s_{min}$,
$s_{max}$ and $|S|$ respectively (c.f.,
Fig.\ref{figure:patient_compare} ). Heuristically this effect can be
seen by observing the range of values seen in
Fig. \ref{figure:patient_pdf_compare_b} versus
Fig. \ref{figure:patient_pdf_compare_a} --- the population of $20,000$
yields a range of glucose values roughly five times that of $D_7$.

\textbf{Non-TDMI graph-based analysis---}The most sensitive PDF
variation metric, $H_{RA}$ points to a relatively diverse
population. In contrast to the results for $D_7$, the less sensitive
PDF variation metric $V_{\bar{\mathcal{S}}}(p)$, also points to a
heterogeneous population; in particular, $V_{\bar{\mathcal{S}}}(p)$ is
just about twice the $V_{\bar{\mathcal{S}}}(p)$ for $D_7$.


\subsubsection{Analysis of the TDMI under variation of $\delta t$}
\label{sec:dynamic_TDMI}

\begin{figure}[tbp]
  \subfigure[TDMI for $\bar{I}$ and $\hat{I}$ with $\delta t$ bins of
    six hours for a period $60$ hours for the $100$ patients with the
    most glucose values using both the histogram and KDE PDF
    estimation techniques.] 
    { 
      \epsfig{file=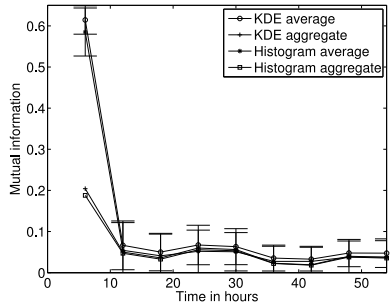,
      height=8cm}
       \label{fig:ehr_ave_v_agg_a}
    }
    \subfigure[TDMI for $\bar{I}$ and $\hat{I}$ with $\delta t$ bins of
    six hours for a period $72$ hours for the $20,000$ randomly selected
    patients using both the histogram and KDE PDF
    estimation techniques.]
    {
      \epsfig{file=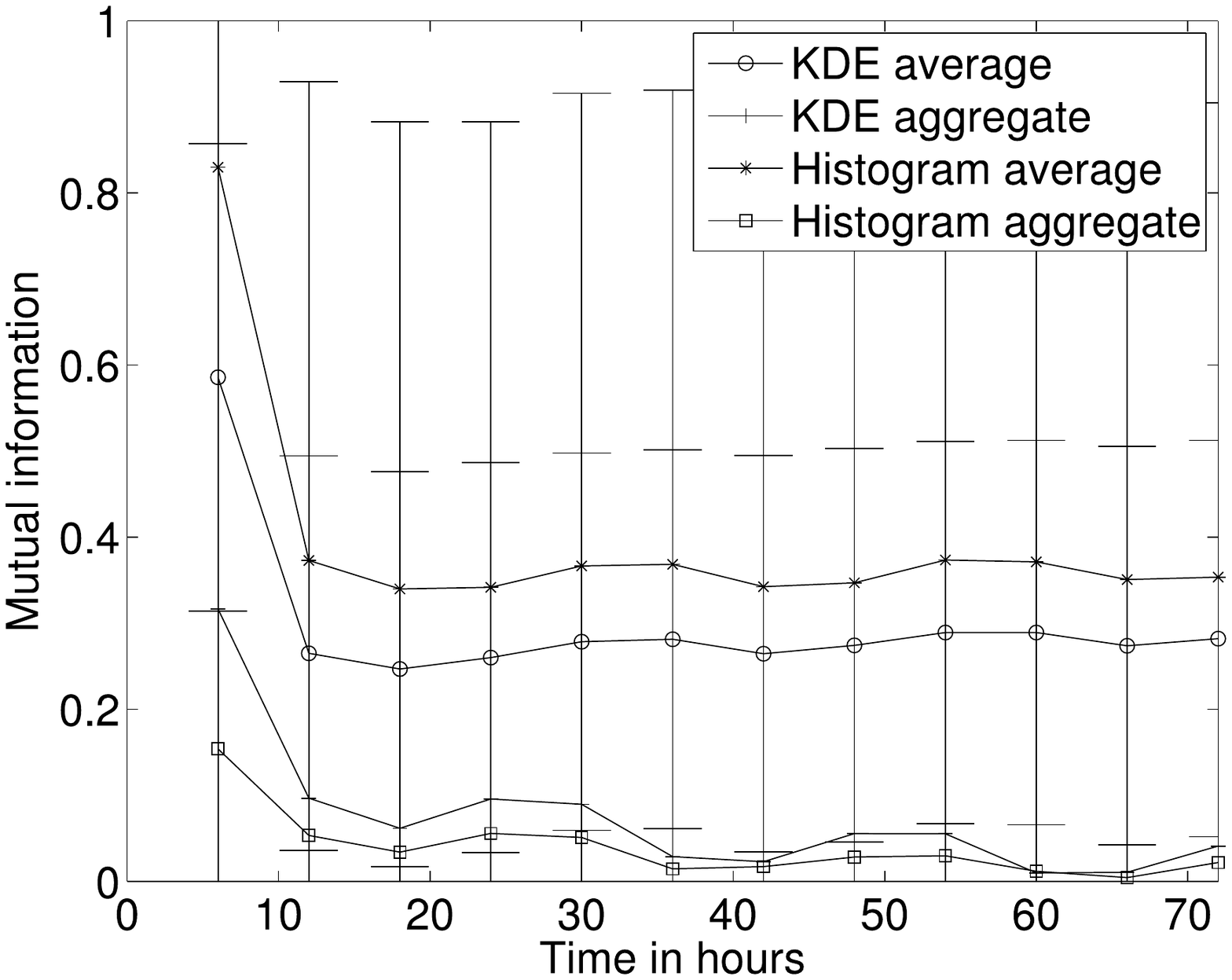, height=8cm}
      \label{fig:ehr_ave_v_agg_b}
    }
    \caption{The TDMI for both $\bar{I}$ and $\hat{I}$ with $\delta t$
      bins of six hours for a period of a few days. With respect to
      Fig. \ref{fig:ehr_ave_v_agg_a} note the following: for $\delta t
      \leq 6 hrs$, $\delta I > 0$; for $\delta t > 6 hrs$, $\delta I
      \approx 0$; the KDE and histogram estimates are extremely
      similar; the diurnal (daily) periodic variation in correlation
      of glucose is clearly evident in both $\bar{I}$ and
      $\hat{I}$. With respect to Fig. \ref{fig:ehr_ave_v_agg_b} note
      the following: for all $\delta t$ $\delta I$ is consistent and
      likely zero within bias; the KDE and histogram estimates differ
      greatly, implying the presence of small sample size effects in
      the average TDMI calculation; the diurnal (daily) periodic
      variation in correlation of glucose is clearly evident in both
      $\bar{I}$ and $\hat{I}$ in all but the KDE estimated TDMI
      average.}  
\label{fig:ehr_ave_v_agg}
\end{figure}

A central motivation for using the TDMI is to observe how nonlinear
correlation evolves in time; however, in the context of a diversely
measured population, one must take care to ensure the TDMI signal
represents a relatively constant population. Relative to $D_7$ and
$D_8$, we know that, for $\delta t$ between $6$ and at least $56$
hours, the representative population is roughly constant. Figure
\ref{fig:ehr_ave_v_agg} details the temporal evolution of the TDMI,
and with it, exhibits five notable features.

First, both data sets display diurnal peaks in predictability; a full
explanation of these peaks, which is dependent the structure of meal
times \cite{pop_phys}. This is scientifically interesting because it
is a signal that can be used to test physiological models, it can be
used to distinguish populations, it implies that outside of very local
time windows, measurements separated by $24$ are more informative than
measurements separated by fewer hours, and finally, the diurnal peaks
confirm the presence of diurnal cycles in humans that are believed to
exist.

Second, relative to $D_7$, the population appears to be
\emph{heterogeneous} on time scales of $6$ hours and less, and
\emph{homogeneous} on time scales longer than $6$ hours. This can be
seen in Fig. \ref{fig:ehr_ave_v_agg_a}, where $\delta I(6hrs)$ is
relatively large and drops to zero by $\delta t =12$ hrs. This is an
interesting result that we are still working to understand.

Third, by comparing the results for $D_7$ and $D_8$, we can observe a
difference in the degree of homogeneity amongst the population. In
particular combining the facts that the error bars for $\bar{I}$ are
large for $D_8$ compared to $D_7$, $\delta I$ is independent of
$\delta t$ for $D_8$, $\delta I$ for $D_8$ is much larger than for
$D_7$, and the broad qualitative TDMI signal (i.e., the diurnal
peaks) is the same for both $D_7$ and $D_8$, it seems clear that both
data sets have somewhat homogeneous populations (i.e., homogeneous
enough to resolve a similar signal), but $D_7$ is considerably more
homogeneous than $D_8$.

Fourth, considering Fig. \ref{fig:ehr_ave_v_agg_b}, it is clear that
the aggregate TDMI resolves the diurnal peaks considerably better
than the average TDMI. This is confirms the usefulness of the
aggregate TDMI in the context of a complex, diversely measured
population.
 
And fifth, the small sample size effects are clearly evident when
comparing the difference between the histogram and KDE estimates of
the TDMI between Figs. \ref{fig:ehr_ave_v_agg_a} and
\ref{fig:ehr_ave_v_agg_b}. In particular, the two different estimates
for the aggregate TDMI on $D_7$ are essentially identical, while the
aggregated TDMI estimates on $D_8$ differ in a nontrivial way (by
more than a factor of two). The average TDMI calculations display an even
stronger effect. Finally, the error bars for $D_8$ are about ten times
the magnitude of the error bars for $D_7$.

The point is, the time evolution of the TDMI is both scientifically
valuable in that it leads to insights not otherwise observed and
interpretable in the context of a time dependent, complex, diversely
measured population using the infrastructure presented in this paper.
 
\subsubsection{Independent analysis of the population composition of
  $D_7$ and $D_8$}
\label{sec:billing}


Based on the time-based information theoretic analysis we have reached
the following population-composition hypotheses: data set seven
represents a homogeneous population for $\delta t > 6 hrs$ and is
heterogeneous for $\delta t \leq 6 hrs$; the subpopulation of data set
eight used to estimate $\hat{I}(\delta t \leq 48)$ is relatively
homogeneous, but less homogeneous than data set seven; overall, data
set $8$ is heterogeneous. However, because these populations are real
patients from a hospital, we can also examine other sources of
information regarding the qualitative types these populations
represent. Specifically, we can consider the
billing codes, which can act as a proxy for population composition,
assigned to the patients in the various populations. It is important
to note that the billing codes are \emph{largely independent of the
  specific lab values}, and thus, can be seen as an outside test of
the validity of the TDMI analysis.

We consider the fraction of patients with the two most frequent
billing codes for three data sets, $D_7$, $D_8$, and the subset of
$D_8$ used to estimate the TDMI-based diagnostics, $D_8'$ (members of
the $D_8'$ subpopulation have at least $10$ glucose measurements
separated by six hours or less). Note that a patient is counted for
having an billing code if it occurs only once. There are two features
of that are important to pay attention to: (i) the overall fraction of
patients that have a given billing code, and (ii), the drop off
between the fraction of patients with the most and second most common
billing codes. For $D_7$, $75 \%$ of the patients are covered by a
single billing code and the drop between the most and second most
common billing codes is around $5 \%$ --- thus $70 + \%$ of these
patients likely have relatively similar afflictions. In contrast, the
most frequently seen billing code in $D_8$ only covers $25 \%$ of the
population, followed by a $10$ point drop off. In constrast, at least
$50 \%$ of $D_8'$ is covered by a single billing code, while the
second most common billing code only covers only a quarter of the
population --- a $25$ point drop. This implies more homogeneity than
$D_8$ but less than $D_7$. Broadly speaking, the billing code analysis
corroborates the conclusions drawn from the time-based information
theoretic analysis in the previous section. Nevertheless, the billing
code analysis, being static, does not reveal the heterogeneity
observed in $D_7$ at $\delta t = 6 hrs$.

\section{Summary}

\textbf{Note, a explicit prescription for interpreting $I$ for a fixed
  time separation $\delta t$ for a population can be found in
  Fig. \ref{fig:TDMI_analysis_schematic} within section
  \ref{sec:interpretation_summary}}. Moreover, an algorithmic
portrayal can be found in appendix \ref{sec:app_pc}.

\textbf{Results of the interpretative framework relative to real
  data.} The methods in this paper were shown to work for both a well
understood computer-generated data set and for a pathologically
diverse real data set. Thus, given a population of time-series that
are: non-uniformly measured in time, of diverse lengths, from
statistically diverse sources, nonstationary, and pathologically
sparse, our methods will likely still yield interpretable results. The
entropy for all populations registered the populations as
diverse. Nevertheless, the TDMI produced a more nuanced picture. In
particular, for one set of patients, the TDMI calculation implied that
a set of patients have differing predictive information up to $6$
hours, and are homogeneous in correlation afterwards.  In contrast,
the same calculation on a heavily filtered general population (the
population that had frequent data measurements), yielded a population
that seemed homogeneous with respect to time-dependent
correlation. While these two sets of patients, according to their
billing codes were similar, they differed in some key features. Thus,
while it is likely that these populations are different, a full
explanation, which requires more clinical study, is beyond the scope
of this paper. Nevertheless, the TDMI analysis yielded results that
were understandable, given this pathologically difficult population of
data.

\textbf{How our method addresses nonstationarity.} At various points
in this paper we have alluded to how nonstationarity is addressed
within our framework. To be more explicit, consider three cases: (i) a
single nonstationary source, (ii) multiple different stationary
sources, and (iii) multiple different nonstationary sources. Relative
to case (i), because there is no real sense of population average,
$\delta I \approx 0$, $\hat{B}_{IRP}=\hat{B}_{PRP}$ and $\mathcal{H}_S
\approx B_E$ --- thus there will be no distinction between
stationarity and nonstationarity. Case (ii) is the case we handled in
section \ref{sec:quant_simulated_data} and does not need
explanation. And case (iii) will behave identically to case (ii);
nonstationary will be difficult to detect, but multiple different
statistical states will be detectable. While it might be too much to
ask to be able to distinguish nonstationarity amongst a population
from a population with multiple stationary sources, we can detect
nonstationarity within an individual, given enough data points. In
particular, relative to case (i), the reason why all the diagnostics
fail to detect multiple statistical states is that there is \emph{no
  concept of averaging over a population}. To address this issue, one
only needs to partition the single time series into multiple pieces
(of sufficient length), and then apply the standard TDMI analysis from
this paper to the new ``population'' of time series. Said differently,
the to detect nonstationarity in a single source, one only needs to
treat the single source as multiple sources and apply our machinery;
if it appears that there are multiple sources, then you know that the
single source has multiple statistical states, and is thus
nonstationary.

\textbf{Comments regarding the connection between the supports and the
  normalizations of the distributions.} In a sense, all support-based
variation amongst the population could be eliminated by normalizing
all individuals to some standard support (or to a distribution with
mean zero and variance one). We did not implement this because
sometimes the normalization of the support matters with respect to the
composition of the population, and we wanted to allow for the TDMI
infrastructure to capture this type of dependence.  Relative to the
example in this paper, having glucose oscillate around $500$ means the
patient is very sick, whereas glucose oscillation around $100$ means
the patient is likely healthy (at least from a blood glucose
perspective) --- we wanted to be able to capture this type of
heterogeneity. That said, if one begins with a \emph{normalized
  population} and performs the TDMI analysis, any $\delta I$
\emph{must} exist because of variation in the \emph{graphs} of the
PDFs. However, if one has enough points per patient to estimate
$\bar{I}$, one knows this anyway upon calculating $\hat{B}_{IRP}$ and
$\hat{B}_{PRP}$; when there are not enough points to estimate $I$ for
every individual, then deducing temporal, graph-based variation is
difficult.

\textbf{Future directions regarding the use of this technique.} One of
the sources of motivation for performing this calculation is based on
the idea of stratifying or clustering populations of individuals by their
predictive information. Based on the TDMI infrastructure here, we have
identified at least $3$ different subpopulations based on their
predictive information structure. Thus future computational problems
will involve developing and testing a more automated form of this
interpretive structure that can be used for generating hypothesized
sub-categories of individuals and eventually an infrastructure that can
be integrated with classification and clustering schemes.

\textbf{Some remaining statistical problems.} In this work we
attempted to outline and show, mathematically, how to interpret the
TDMI and information entropy for aggregated populations. Nevertheless,
there are many details that are remain. In particular, a partial list
might include full rigorous proofs regarding: the technical conditions
under which our claims (i.e., $\delta I >0$ if an only if $\epsilon_i
>0$ for some $i$) apply; the convergence properties of various
quantities we propose (i.e., $\delta I$, $\mathcal{H}_S$. etc); and
the full relationships between what the information entropy and TDMI
can imply about one another. The goal of this work was to propose a
practically workable framework calculating the TDMI for complex
populations of time series. However, this work leaves many
interesting, more abstract questions remaining.

\section{Acknowledgments}

The authors would like to thank J. Dias, N. Elhadad, A. Perotte, and
D. Varn for carefully reading this paper and providing many useful
comments.  DJA would like to thank C. Shalizi for early discussions
related to this work.  Finally, the authors would like to acknowledge
the financial support provided by NLM grant RO1 LM06910.

\appendix

\section{Analysis of aggregation order}
\label{sec:app1}

\subsection{Detailed average TDMI calculation}
\label{sec:app1.1}

Begin by recalling the definition of the average TDMI:
\begin{align}
\label{equation:ibar_app}
&\bar{I}(\tau) = \frac{1}{N} \sum_{i=1}^N \int p(X_i(j),
  X_i(j-\tau)) \\ \notag
&\log
  (\frac{p(X_i(j),X_i(j-\tau))}{p(X_i(j))p(X_i(j-\tau))} ) dX_i(t)
  dX_i(t+\tau) \\ \notag
&= \int \bar{\iota}(\tau) dX(t) dX(t+\tau).
\end{align}
Next, recall that for the average TDMI, we have PDFs defined entirely
with respect to the abstract support, $\bar{\mathcal{S}}$. In this
situation, we define the $i^{th}$ PDF relative to the ``average'' PDF,
$p_1$, by:
\begin{equation}
\label{equation:p_1_bar}
p_i = p_1(\bar{\mathcal{S}}) - \bar{\epsilon}_i(\bar{\mathcal{S}})
\end{equation}
where $\bar{\epsilon}_i(\bar{\mathcal{S}})$ is distance between the
\emph{graphs} of $p_1$ and $p_i$ at a given value in
$\bar{\mathcal{S}}$. Next, for convenience, define the following:
$p(X_i(j), X_i(j-\tau)) = p(j,\tau)$, $p(X_i(j))=p(j)$,
$p(X_i(j-\tau))=p(\tau)$, $\bar{\epsilon}_i(\bar{\mathcal{S}}) =
\bar{\epsilon}_i$, $p_i(j,\tau) = p_1(j,\tau) - \bar{\epsilon}_i$, $p_i(j) =
p_1(j) - \bar{\epsilon}_i$, and $p_i(\tau) = p_1(\tau) - \bar{\epsilon}_i$. With
this notation, we can now re-write \emph{the integrand} in
Eq. \ref{equation:ibar_app}
\begin{align}
=& \frac{1}{N} [ p_1(j,\tau) \log(\frac{p_1(j,\tau)}{p_1(j) p_1(\tau)}
)+ \\ 
&\sum_{i=2}^N (p_1(j,\tau) - \bar{\epsilon}_i) \log(\frac{p_1(j,\tau) -\bar{\epsilon}_i}{(p_1(j)-\bar{\epsilon}_i)( p_1(\tau)-\bar{\epsilon}_i) }) ]
\end{align}
Next, factoring $\frac{p_1(j,\tau)}{p_1(j)p_1(\tau)}$ out of the
summation term, one arrives at: 
\begin{align}
=& \frac{1}{N} [ p_1(j,\tau) \log(\frac{p_1(j,\tau)}{p_1(j) p_1(\tau)}
)+ \\ 
&\sum_{i=2}^N (p_1(j,\tau) - \bar{\epsilon}_i) [
\log(\frac{p_1(j,\tau)}{(p_1(j))( p_1(\tau)) }) + \\
 & \log(\frac{1-\frac{\bar{\epsilon}_i}{p_1(j,\tau)}}{ 1 -
  \frac{\bar{\epsilon}_i}{p_1(j)p_1(\tau)} ( p_1(j) + p_1(\tau)) + \frac{\bar{\epsilon}_i^2}{p_1(j)p_1(\tau)}  }) ] ].
\end{align}
Multiplying and collecting terms under the sum, one obtains:
\begin{align}
=& \frac{1}{N} [N p_1(j,\tau) \log(\frac{p_1(j,\tau)}{p_1(j) p_1(\tau)}
)+ \\ 
&\sum_{i=2}^N \bar{\epsilon}_i [
\log(\frac{p_1(j,\tau)}{(p_1(j))( p_1(\tau)) }) + \\
 & \log(\frac{1-\frac{\bar{\epsilon}_i}{p_1(j,\tau)}}{ 1 -
  \frac{\bar{\epsilon}_i}{p_1(j)p_1(\tau)} ( p_1(j) + p_1(\tau)) +
  \frac{\bar{\epsilon}_i^2}{p_1(j)p_1(\tau)}  }) ] + \\ 
& p_1(j,\tau) \log(\frac{1-\frac{\bar{\epsilon}_i}{p_1(j,\tau)}}{ 1 -
  \frac{\bar{\epsilon}_i}{p_1(j)p_1(\tau)} ( p_1(j) + p_1(\tau)) +
  \frac{\bar{\epsilon}_i^2}{p_1(j)p_1(\tau)}  }) ] \\
=& \bar{\rho}(\tau) + \frac{1}{N} [\sum_{i=2}^N \bar{\epsilon}_i [
\log(\frac{p_1(j,\tau)}{(p_1(j))( p_1(\tau)) }) + \\
 & \log(\frac{1-\frac{\bar{\epsilon}_i}{p_1(j,\tau)}}{ 1 -
  \frac{\bar{\epsilon}_i}{p_1(j)p_1(\tau)} ( p_1(j) + p_1(\tau)) +
  \frac{\bar{\epsilon}_i^2}{p_1(j)p_1(\tau)}  }) ] + \\ 
& p_1(j,\tau) \log(\frac{1-\frac{\bar{\epsilon}_i}{p_1(j,\tau)}}{ 1 -
  \frac{\bar{\epsilon}_i}{p_1(j)p_1(\tau)} ( p_1(j) + p_1(\tau)) +
  \frac{\bar{\epsilon}_i^2}{p_1(j)p_1(\tau)}  }) ] \\
=& \bar{\rho}(\tau) + \bar{G}(\tau)
\end{align}
where $\bar{G}(\tau)$ can be shown to have the more digestible form:
\begin{equation}
\label{equation:gbar_app}
\begin{split}
\bar{G}(\tau) &= \\
&- \frac{1}{N} [ \sum_{i=1}^{N-1}
\left( \frac{\bar{\epsilon}_i}{p(X_1(j), X_1(j-\tau))} \right) \\
&\left(\log \frac{p(X_1(j),
  X_1(j-\tau))}{p(X_1(j))p( X_1(j-\tau))} \right) \\
&+ \log \left(
\frac{1-\frac{\bar{\epsilon}_i}{p(X_1(j),X_1(j-\tau))}}{(1-\frac{\bar{\epsilon}_i}{p(X_1(j))})
  (1-\frac{\bar{\epsilon}_i}{p(X_1(j-\tau))} )} \right) \\
&\left( \frac{\bar{\epsilon}_i}{p(X_1(j), X_1(j-\tau))} - 1 \right) ]
\end{split}
\end{equation}

\subsection{Detailed aggregate TDMI calculation}
\label{sec:app1.2}

Begin by recalling the definition of the TDMI for an aggregate population:
\begin{align}
\label{equation:aggregated_MI_def_app}
\hat{I}(\tau) =& \int p(X_1^{n-\tau}; X_{\tau}^{n}) \log
 (\frac{p(X_1^{n-\tau}; X_{\tau}^{n})}{p(X_1^{n-\tau})
   p(X_{\tau}^{n})})dX_1^{n-\tau} dX_{\tau}^{n} \\ \notag
=& \int \hat{\iota}(\tau) dX_1^{n-\tau} dX_{\tau}^{n}
\end{align}
Next, recall that in this situation we first select an ``average'' PDF
relative to the abstract support $\hat{\mathcal{S}}$ and then we
define the $i^{th}$ PDF relative to this ``average'' PDF on the
\emph{total support} $\hat{S}$, $p_1$, by:
\begin{equation}
\label{equation:p_1_hat_app}
p_i = p_1(\hat{S}) - \hat{\epsilon}_i(\hat{S})
\end{equation}
where $\hat{\epsilon}_i(\hat{S})$ is distance between the
\emph{graphs} of $p_1$ and $p_i$ at a given value in
$\hat{S}$. Next, for convenience, define the following:
$p(X_i(j), X_i(j-\tau)) = p_i(j,\tau)$, $p(X_i(j))=p_i(j)$,
$p(X_i(j-\tau))=p_i(\tau)$, $\hat{\epsilon}_i(\hat{S}) =
\hat{\epsilon}_i$, $p_i(j,\tau) = p_1(j,\tau) - \hat{\epsilon}_i$, $p_i(j) =
p_1(j) - \hat{\epsilon}_i$, and $p_i(\tau) = p_1(\tau) -
\hat{\epsilon}_i$, never forgetting that all of these quantities
depend on a particular value in the support, $\hat{S}$. With
this notation, we can now re-write \emph{the integrand} in
Eq. \ref{equation:aggregated_MI_def_app} in terms of only $p_1$ and
$\hat{\epsilon}$, arriving at:
\begin{align}
=&\frac{1}{N} \sum_{i=1}^N (p_1(j,\tau) - \hat{\epsilon}_i) \\
&(\log(
\frac{\frac{1}{N} \sum_{i=1}^N (p_1(j,\tau) - \hat{\epsilon}_i)}{
  (\frac{1}{N} \sum_{i=1}^N (p_1(j) - \hat{\epsilon}_1))
(\frac{1}{N} \sum_{i=1}^N (p_1(\tau) - \hat{\epsilon}_1))}))
\end{align}
Next, factoring $p_1(j,\tau)$, $p_1(j)$, and $p_1(\tau)$ out of the
numerator of the summation terms, one arrives at: 
\begin{align}
=& (\frac{p_1(j,\tau)}{N} \sum_{i=1}^N (1 -
\frac{\hat{\epsilon}_i}{p_1(j,\tau)})) \\
& \log(  \frac{\frac{p_1(j,\tau)}{N} \sum_{i=1}^N (1 -
\frac{\hat{\epsilon}_i}{p_1(j,\tau)}) } { (\frac{p_1(j)}{N} \sum_{i=1}^N (1 -
\frac{\hat{\epsilon}_i}{p_1(j)}))  (\frac{p_1(\tau)}{N} \sum_{i=1}^N (1 -
\frac{\hat{\epsilon}_i}{p_1(\tau)})))}
\end{align}
which, after collecting terms, becomes:
\begin{align}
\hat{\iota} =& (p_1(j,\tau) - \sum_{i=1}^N \frac{\hat{\epsilon}}{N
  p_1(j,\tau)}) \\
&( \log(\frac{p_1(j,\tau)}{p_1(j)p_1(\tau)}) + \log \frac{ 1-\sum_{i=1}^N
  \frac{\hat{\epsilon}_i}{Np_1(j,\tau)} }{(1-\sum_{i=1}^N
  \frac{\hat{\epsilon}_i}{Np_1(j)}) ( 1-\sum_{i=1}^N
  \frac{\hat{\epsilon}_i}{Np_1(\tau)}) }).
\end{align}
Next, collecting the $p_1(j,\tau)\log(\frac{p_1(j,\tau)}{p_1(j)p_1(\tau)})$ term,
one gets:
\begin{align}
\hat{\iota} =p_1(j,\tau)\log(\frac{p_1(j,\tau)}{p_1(j)p(\tau)}) + \hat{G}(\tau) 
\end{align}
where $\hat{G}$ is given by:
\begin{equation} 
\label{equation:ghat_app}
\begin{split}
\hat{G}(\tau) =& \log \left(  \frac{1 - \frac{\sum_{i=1}^{N-1}
    \hat{\epsilon}_i}{N p_1(j,\tau)}}{(1 -
  \frac{\sum_{i=1}^{N-1} \hat{\epsilon}_i}{N p_1(j)})( 1 - \frac{\sum_{i=1}^{N-1}
    \hat{\epsilon}_i}{N p_1(\tau)})}   \right) \\
&\left(
p_1(j, \tau) - \frac{\sum_{i=1}^{N-1}
  \hat{\epsilon}_i}{N}  \right) \\
-& \frac{\sum_{i-1}^{N-1}\hat{\epsilon}_i}{N} \log \left(  \frac{p_1(j, \tau)}{p_1(j)
   p_1(\tau)} \right)
\end{split}
\end{equation}

\subsection{Pseudocode for interpreting the TDMI for a population of
  time series}
\label{sec:app_pc}

\begin{algorithm}
\caption{How to interpret the TDMI for a population of time series}
\label{algorithm:tdmi}
\begin{algorithmic}
\IF{there are enough points to estimate $\bar{I}$ (usually $\sim 100$
  \emph{pairs} of points \emph{per representative individual} are required)}
\STATE estimate $\delta I$ and $H_{\Theta}$

\IF{$\delta I > B_{IRP}$}
\STATE the population is heterogeneous

\IF{$\mathcal{H}_S \sim 0$}
\STATE supports (or ranges) are diverse or disjoint
\ELSIF{$\mathcal{H}_S \sim 1$}
\STATE supports (or ranges) are uniform
\ENDIF

\ELSIF{$\delta I \leq B_{IRP}$}
\STATE the population is homogeneous
\ENDIF

\IF{$H_{\Theta} \sim 0$}
\STATE  the population is well represented
\ELSIF{$H_{\Theta} \sim 1$}
\STATE the portions of the population are overrepresented
\ENDIF

\ELSIF{not enough pairs to estimate $\bar{I}$}
 \STATE estimate $\hat{I}$,  $\mathcal{H}_S$, and $H_{\Theta}$

\IF{$\mathcal{H}_S \sim 0$}
\STATE supports (or ranges) are diverse or disjoint

\IF{there are enough pairs of points per patient to estimate a PDF for
each patient at the specific $\delta t$}
\STATE $V_{\hat{S}}(p)$ (i.e., $V(p)$ relative to the abstract supports)

\IF{$V_{\hat{S}}(p) \sim 1$}
\STATE the population used to estimate $\hat{I}$ has graph-based heterogeneity
\ELSIF{$V_{\hat{S}}(p) \sim 0$}
\STATE the population used to estimate $\hat{I}$ is graphically homogeneous

\ENDIF

\ELSIF{it is not possible to accurately estimate a PDF for each
  patient at the specific $\delta t$}
\STATE it is not possible to determine the contribution of the
graph-based heterogeneity to the overall heterogeneity

\ENDIF

\ELSIF{$\mathcal{H}_S \sim 1$}
\STATE supports (or ranges) are uniform

\IF{$V_{\bar{\mathcal{S}}}(p) \sim 1$}
\STATE the population used to estimate $\hat{I}$ has graph-based heterogeneity
\ELSIF{$V_{\bar{\mathcal{S}}}(p) \sim 0$}
\STATE the population used to estimate $\hat{I}$ is homogeneous
\ENDIF

\ENDIF

\IF{$H_{\Theta} \sim 0$}
\STATE  the population is well represented
\ELSIF{$H_{\Theta} \sim 1$}
\STATE the portions of the population are overrepresented
\ENDIF

\ENDIF

\COMMENT{NOTE: there are $10$ possible sharp interpretations for
  \emph{both} $\delta I$ and $\hat{I}$-only cases.}

\COMMENT{All TDMI interpretations should include: $I$-like quantities
  (e.g., $\hat{I}$, $\delta I$, etc), population diversity
  qualification (support- and graph-based contributions to diversity;
  if they are unknown, this should be specified), and the make-up of
  the population used to estimate the $I$-based quantities (e.g., $H_{\Theta}$.}

\COMMENT{NOTE: even under the best circumstances, it may be difficult to determine what proportion of
  the heterogeneity is due to support-based versus graph-based
  diversity.}

\end{algorithmic}
\end{algorithm}


\end{document}